\RequirePackage{amsmath}
\documentclass[sw]{iosart2x}

\usepackage{graphicx}
\usepackage{dirtytalk}
\usepackage{ragged2e}
\usepackage{wasysym}
\usepackage{color}
\usepackage[table,xcdraw]{xcolor}
\usepackage{lscape}
\usepackage{longtable}
\usepackage{soul}
\usepackage{enumitem}
\usepackage{hyperref}
\usepackage{url,footnote}
\usepackage{dirtytalk}
\usepackage{textcomp}
\usepackage[utf8]{inputenc}
\usepackage[T1]{fontenc}
\usepackage[english]{babel}
\usepackage{sidecap}
\usepackage{subcaption}
\usepackage{wrapfig}
\usepackage{multirow}
\usepackage{listings}
\usepackage{amsmath}
\usepackage{stmaryrd}
\usepackage{amssymb}
\usepackage[linesnumbered,ruled,lined,boxed,commentsnumbered]{algorithm2e}

\SetCommentSty{mycommfont}
\usepackage{ifsym}
\usepackage{booktabs}
\usepackage{xparse}   
\NewDocumentCommand{\rot}{O{90} O{1em} m}{\makebox[#2][l]{\rotatebox{#1}{#3}}}%

\theoremstyle{definition}
\newtheorem{definition}{Definition}[section]

\definecolor{codegreen}{rgb}{0,0.6,0}
\definecolor{codegray}{rgb}{0.5,0.5,0.5}
\definecolor{codepurple}{rgb}{0.58,0,0.82}
\definecolor{backcolour}{rgb}{0.95,0.95,0.92}
 
\lstdefinestyle{mystyle}{
    backgroundcolor=\color{backcolour},   
    commentstyle=\color{codegreen},
    keywordstyle=\color{magenta},
    numberstyle=\tiny\color{codegray},
    stringstyle=\color{codepurple},
    basicstyle=\footnotesize,
    breakatwhitespace=false,         
    breaklines=true,                 
    captionpos=b,                    
    keepspaces=true,                   
    numbers=left,                    
    numbersep=2pt,                  
    showspaces=false,                
    showstringspaces=false,
    showtabs=false,                  
    tabsize=1
}
 
\hypersetup{
    colorlinks=true,
    linkcolor=red,
    filecolor=blue,      
    urlcolor=cyan,
}
\usepackage[textsize=small]{todonotes} 
\presetkeys%
    {todonotes}%
    {inline}{}

\pubyear{0000}
\volume{0}
\firstpage{1}
\lastpage{1}

\begin{document}

\begin{frontmatter}

\title{A Stitch in Time Saves Nine -- SPARQL Querying of Property Graphs using Gremlin Traversals}


\runningtitle{Gremlinator}


\author[A]{\inits{N.}\fnms{Harsh} \snm{Thakkar}\ead[label=e1]{\{thakkar, jens.lehmann\}@cs.uni-bonn.de}%
\thanks{Corresponding author. \printead{e1}.}},
\author[B]{\inits{N.N.}\fnms{Dharmen} \snm{Punjani}\ead[label=e2]{dpunjani@di.uoa.gr}}
\author[C]{\inits{N.N.}\fnms{Yashwant} \snm{Keswani}\ead[label=e3]{201301047@daiict.ac.in}}
\author[A,D]{\inits{N.N.}\fnms{Jens} \snm{Lehmann}\ead[label=e4]{jens.lehmann@iais.fraunhofer.de}}
and
\author[E]{\inits{N.N.}\fnms{S{\"o}ren} \snm{Auer}\ead[label=e5]{soeren.auer@tib.eu}}
\runningauthor{H. Thakkar et al.}
\address[A]{Smart Data Analytics, \orgname{University of Bonn}, \cny{Germany}\printead[presep={,\hspace{5pt} }]{e1}}
\address[B]{Department, \orgname{National and Kapodistrian University of Athens}, \cny{Greece}\printead[presep={,\hspace{5pt}}]{e2}}
\address[C]{\orgname{DA-IICT}, \cny{India}\printead[presep={,\hspace{5pt}}]{e3}}
\address[D]{\orgname{Fraunhofer IAIS}, \cny{Germany}\printead[presep={,\hspace{5pt}}]{e3}}
\address[E]{TIB Technische Informationsbibliothek \& L3S Research Center, \orgname{Leibniz University of Hannover}, \cny{Germany}\printead[presep={,\hspace{5pt}}]{e5}}


\begin{abstract}
Knowledge graphs have become popular over the past years and frequently rely on the Resource Description Framework (RDF) or Property Graphs (PG) as underlying data models. 
However, the query languages for these two data models -- SPARQL for RDF and Gremlin for property graph traversal -- are lacking interoperability. 
We present Gremlinator, a novel SPARQL to Gremlin translator. 
Gremlinator translates SPARQL queries to Gremlin traversals for executing graph pattern matching queries over graph databases. 
This allows to access and query a wide variety of Graph Data Management Systems (DMS) using the W3C standardized SPARQL query language and avoid the learning curve of a new Graph Query Language.
Gremlin is a system agnostic traversal language covering both OLTP graph database or OLAP graph processors, thus making it a desirable choice for supporting interoperability wrt. querying Graph DMSs. 
We present a comprehensive empirical evaluation 
of Gremlinator and demonstrate its validity and applicability by executing SPARQL queries on top of the leading graph stores Neo4J, Sparksee and Apache TinkerGraph and compare the performance with the RDF stores Virtuoso, 4Store and JenaTDB. 
Our evaluation demonstrates the substantial performance gain obtained by the Gremlin counterparts of the SPARQL queries, especially for star-shaped and complex queries.
\end{abstract}


\end{frontmatter}


\section{Introduction}


Knowledge graphs have become increasingly popular over the past years. 
The two most popular data models for representing and storing knowledge graphs are property graphs (PG) and the Resource Description Framework (RDF). 
For RDF, the SPARQL query language was standardized by W3C, whereas for PGs several languages are frequently used, including Gremlin~\cite{rodriguez2015traversalmachine}.
Both data models and the corresponding data management techniques have distinct and complementary characteristics: RDF is suited for distributed data integration with built-in world-wide unique identifiers and the expressive SPARQL query language; PGs on the other hand support extremely scalable storage and querying and are meanwhile widely used for modern Web applications.

In this article, we present an approach for executing SPARQL queries over graph databases via Gremlin traversals -- \textit{Gremlinator}, thus building a bridge between the currently still largely disjoint semantic and graph data technology ecosystems and thus addressing the query interoperability problem.

A SPARQL-PG query translation renders several benefits: 
(1) Applications based on W3C Semantic Web standards, like SPARQL and RDF, can use property graph databases in a non-intrusive fashion.
(2) The query translation lays the foundation for a hybrid use of RDF triple stores and property graph DMS -- where a particular query can be dispatched to the DMS capable to answer the query more efficiently~\cite{das2014tale}.
In particular, property graph databases have been shown to work very well for a wide range of queries which benefit from locality in a graph. 
Rather than performing expensive joins, property graph databases use micro indices to perform traversals. 
(3) Users familiar with the W3C SPARQL query language can avoid learning another query language.

To the best of our knowledge, this is the first work addressing the query interoperability (translation) problem. 
Related work (cf. Section~\ref{sec:relwork}) mostly covers the area of SPARQL to SQL conversion and vice versa. 
In contrast to those previous efforts, we have to overcome the challenge of mediating between two very different execution paradigms: 
While SPARQL uses pattern matching techniques, Gremlin is based on performing graph traversals.
More specifically, previous efforts applied query rewriting techniques between formalisms, which are ultimately rooted in relational algebra operations, whereas we had to bridge more disparate query paradigms. 
While this is a significant challenge, it is also the reason why substantial performance improvements can be made depending on the query characteristics: 
Whereas direct SPARQL query execution can be expected to be suitable for large analytical joins over the entire dataset, the Gremlin conversion can significantly speed up all queries that require exploiting the graph locality.

\begin{figure}[htbp]
\begin{center}
\includegraphics[width=0.5\textwidth]{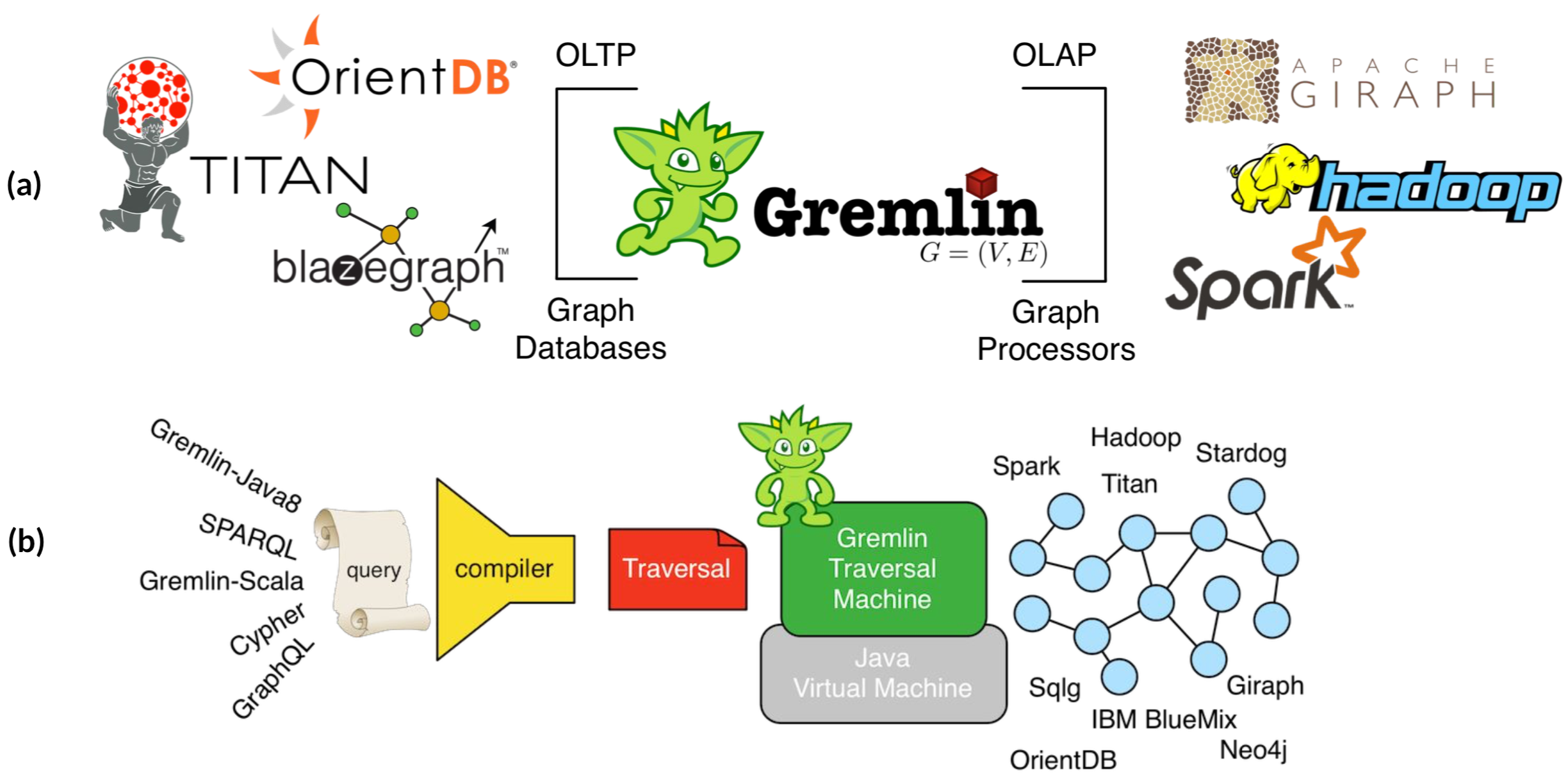}
\end{center} \vspace{-5pt}
\caption{\textbf{The Gremlin Traversal Language and Machine.}}
\label{fig:tinker-all}
\end{figure}

We selected \textit{TinkerPop Gremlin} as target language, since it is more general than, e.g.~CYPHER, as it supported by a wide range of property graph databases (including OLTP and OLAP processors (see Figure~\ref{fig:tinker-all} (a)). 
Moreover, Gremlin supports both the imperative (graph traversal) and declarative (graph pattern matching) style~\cite{rodriguez2015traversalmachine}, for addressing the query interoperability issue.
Lastly, together with the \emph{Apache TinkerPop} framework, Gremlin is a language and a virtual machine, enabling to design another traversal language that compiles to the Gremlin traversal machine (analogous to how Scala compiles to the JVM), ref. Figure~\ref{fig:tinker-all} (b). 

We map SPARQL queries to the pattern matching Gremlin traversals (i.e. we map declarative SPARQL queries to declarative Gremlin constructs and not the imperative ones).
This ensures a level of fairness when comparing the performance of both Graph Query Languages (GQLs). 
Furthermore, instead of translating SPARQL queries to a specific dialect of Gremlin (e.g.~Gremlin-Java8, Gremlin-Python etc.), we map each corresponding operation within a SPARQL basic graph pattern (BGP) to its corresponding traversal step in the Gremlin instruction library (i.e. a single step traversal operation).
As a result, we build complex pattern matching traversals, in an analogous fashion to SPARQL style querying wherein multiple BGPs form complex graph patterns (CGP).
Thus, it is possible to construct a corresponding Gremlin traversal for each SPARQL query.

Overall, we make the following contributions:
\begin{itemize}[nosep]
    \item We propose a novel approach for mapping SPARQL queries to Gremlin pattern matching traversals, , which is the first work converting an RDF to a property graph query language to the best of our knowledge.
    \item Our Gremlinator implementation for executing SPARQL queries over a plethora of third party graph DMS such as \textit{Neo4J}, \textit{Sparksee}, \textit{OrientDB}, etc. using the \textit{Apache TinkerPop} framework is openly available.
    \item We report the results of a comprehensive empirical evaluation of the proposed translation approach comprising a variety of state-of-the-art property graph databases and triple stores on the Northwind and BSBM datasets.
\end{itemize}

The remainder of the article is organized as follows:
Section~\ref{sec:relwork} covers related query conversion efforts. 
Section~\ref{sec:prelims} introduces preliminary notions. 
Section~\ref{sec:gpm_gremlin} describes the relationship between SPARQL graph pattern matching and Gremlin traversal steps. 
Section~\ref{sec:approach} explains our mapping approach. 
Section~\ref{sec:evaluation} presents the experimental evaluation on two famous datasets, discusses the results and observations. 
Finally, Section~\ref{sec:conclusion} concludes the article and describes future work.

\section{Related Work}\label{sec:relwork}
In this section we briefly survey the related work with regard to techniques and tools that support the translation and execution of GQLs.

\textbf{SPARQL $\rightarrow$ SQL:} There is a substantial amount of work been done for conversion of SPARQL queries to SQL queries~\cite{calvanese2017ontop,rodriguez2015efficient,elliott2009complete,chebotko2009semantics,zemke2006converting,priyatna2014formalisation}. 
\emph{Ontop}~\cite{calvanese2017ontop}\footnote{Ontop system (\url{http://ontop.inf.unibz.it/})} exposes relational databases as virtual RDF graphs by linking the terms (classes and properties) in the ontology to the data sources through mappings. 
This virtual RDF graph can then be queried using SPARQL by dynamically and transparently translating the SPARQL queries into SQL queries over the relational databases.
The work presented in \cite{rodriguez2015efficient} generates SQL that is optimized and also provides a well-defined specification of the SPARQL semantics used in the translation.
In addition, Ontop also supports R2RML mappings over general relational schemas. 
The authors show that their implementation can outperform other well known SPARQL-to-SQL systems, as well as commercial triple stores by large margin.
In~\cite{elliott2009complete} a SPARQL-to-SQL translation technique is introduced, that focuses on the generation of efficient SQL queries.
It relies on a mapping language that lacks support for URI templates and is less expressive than R2RML.
\cite{chebotko2009semantics} proposes a translation function that takes a query and two many-to-one mappings: 
(i) a mapping between the triples and the tables, and 
(ii) a mapping between pairs of the form (triple, pattern, position) and relational attributes. 
In addition, the approach in~\cite{chebotko2009semantics} assumes that the underlying relational DB is denormalized, and stores RDF terms.
The two semantics deviate in the definition of the \texttt{OPTIONAL} algebra operator.
The work in~\cite{priyatna2014formalisation} is the extension of work in~\cite{chebotko2009semantics} to include R2RML mappings.
\cite{zemke2006converting} makes use of non-standard SQL constructs for SPARQL–SQL translation and lacks the formal proof that the translation is correct and an empirical evaluation with realistic data is missing.

\textbf{SQL $\rightarrow$ SPARQL:} The work in~\cite{rachapalli2011retro} presents a formal semantics preserving the translation from SQL to SPARQL. \emph{RETRO}~\cite{rachapalli2011retro} deals only with schema mapping and query mapping rather than to transform the data physically. 
Schema mapping derives a domain-specific relational schema from RDF data. Query mapping transforms an SQL query over the schema into an equivalent SPARQL query, which in turn is executed against the RDF store.

\textbf{SQL $\rightarrow$ CYPHER:} CYPHER\footnote{CYPHER Query Language~(\url{https://neo4j.com/developer/cypher-query-language/})} is the graph query language used to query the \emph{Neo4j}\footnote{\label{Neo4j}Neo4j~(\url{https://neo4j.com/})} graph database. 
There has been no work yet aiming to convert the RDBMS to CYPHER. 
However, there are some examples\footnote{SQL to CYPHER~(\url{https://neo4j.com/developer/guide-sql-to-cypher/})} that show the  equivalent CYPHER queries for certain SQL queries.

\section{Preliminaries}\label{sec:prelims}
In this section, we recall and summarize the mathematical concepts which will be used in this article. 
Our notation closely follows~\cite{AnglesFMGQLs16} and extends~\cite{DBLP:conf/icde/RodriguezN11} by introducing the notion of vertex labels, a detailed discussion on which can be found in~\cite{thakkar2017graph}. 

\subsection{Graph Data Models}\label{gdms}

\subsubsection{Edge-labeled Graphs.}\label{sec:rdf_graph}
The Resource Description Framework (RDF) is a well-known W3C standard, which is used for data modeling and encoding machine readable content on the Web~\cite{world2014rdf} and within intranets.
An RDF graph can be seen as a set of triples, roughly analogous to nodes and edges in a graph database. 
However, RDF is more specific in defining disjoint vertex-sets of blank nodes, literals and IRIs.
In the rudimentary form, an RDF graph is a directed, edge-labeled, multi-graph or simply an edge-labeled graph.
In our current context, we do not consider blank nodes.\footnote{The treatment of blank nodes is orthogonal to our current goal, as they related to the translation RDF graphs to property graphs. We focus on the pattern matching features and semantics of SPARQL and Gremlin.}
Edge-labeled graphs can be used to encode complex information despite their elementary structure~\cite{AnglesFMGQLs16}.
Edge-labeled graphs have been formally defined in a wide variety of texts, such as~\cite{cyganiak2005relational,perez2006semantics,AnglesFMGQLs16,reutter2013graph,DBLP:journals/corr/NguyenLBS16}. 
We adapt the definition provided by~\cite{AnglesFMGQLs16}, which is the closest to our current context:

\begin{definition}[Edge-labeled Graph]\label{def:rdf}
An edge-labeled graph is defined as $G = \{V, E\}$; where: 
\begin{itemize}[nosep]
    \item $V$ is the set of vertices, 
    \item $E$ is the set of \textit{directed} edges such that $E \subseteq (V \times Lab \times V)$ where $Lab$ is the set of Labels.  \qed
\end{itemize}
\end{definition}

\begin{figure*}[tb]
    \centering
    \includegraphics[scale=0.3]{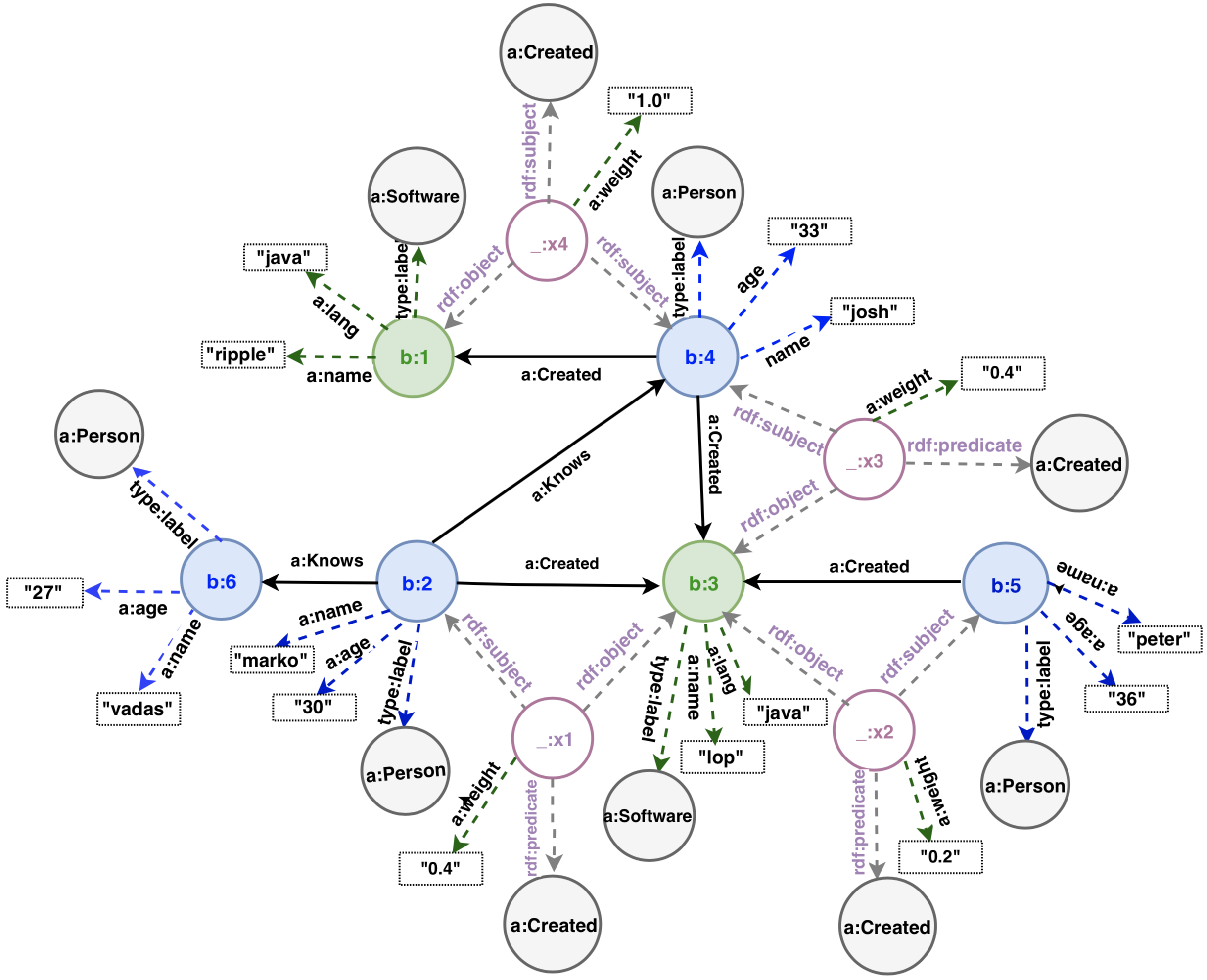}
    \caption{\textbf{RDF graph example.} This figure portrays a collaboration network of employees in a software company.
    }
    \label{fig:rdf_graph}
\end{figure*}
       
\subsubsection{Property Graphs}\label{sec:pg_graph}
Property graphs, also referred to as directed, edge-labeled, attributed multi-graphs, have been formally defined in a wide variety of texts, such as~\cite{AnglesFMGQLs16,ReinventingGubichevT14,krause2016sql,DBLP:books/igi/Sakr11/RodriguezN11,prud2006sparql}. 
We adapt the definition of property graphs presented by~\cite{DBLP:books/igi/Sakr11/RodriguezN11}: 

\begin{definition}[Property Graph]\label{def:pg}
A property graph is defined as $G = \{V, E, \lambda, \mu\}$; where: 
\begin{itemize}[nosep]
    \item $V$ is the set of vertices,
    \item $E$ is the set of \textit{directed} edges such that $E \subseteq (V \times Lab \times V)$ where $Lab$ is the set of Labels, 
    \item $\lambda$ is a function that assigns labels to the edges and vertices (i.e. $\lambda : V \cup E \rightarrow \Sigma^{*}$)\footnote{A finite set of strings ($\Sigma^{*}$)}, and 
    \item $\mu$ is a partial function that maps elements and keys to values (i.e. $\mu : (V \cup E) \times R \rightarrow S)$ i.e. properties (key $r \in R$, value $s \in S$). \qed
\end{itemize} 
\end{definition}

\begin{figure}[tb]
    \centering
    \includegraphics[scale=0.15]{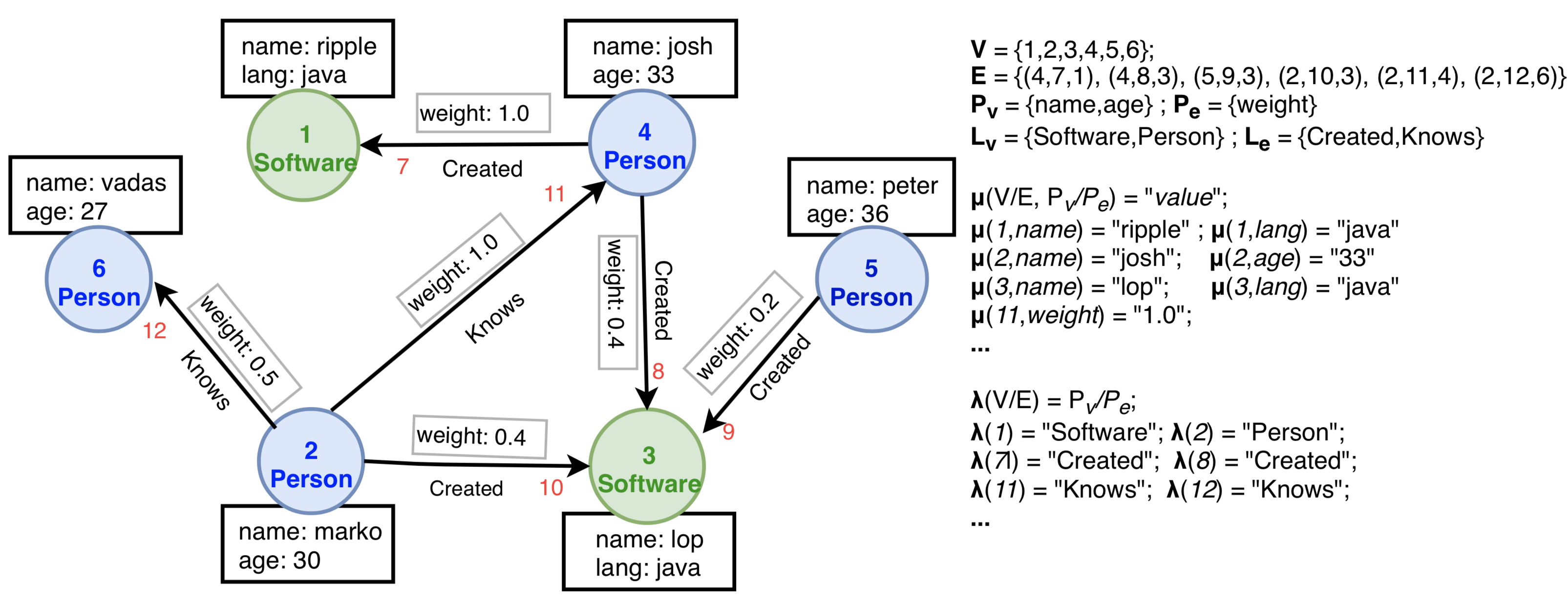} 
    \caption{\textbf{Property Graph example.} This figure presents the property graph version of the RDF graph as in Figure~\ref{fig:rdf_graph}}
    \label{fig:property_graph}
\end{figure}

Figures~\ref{fig:rdf_graph} and~\ref{fig:property_graph}, present different data model visualizations of the \textit{Apache TinkerPop} modern crew graph\footnote{TinkerPop Modern Crew property graph (\url{http://tinkerpop.apache.org/docs/3.2.3/reference/\#intro})}. We use these as running examples throughout this article.

\subsection{Graph Pattern Matching}
The Graph Pattern Matching (GPM) problem is generally perceived as a subgraph matching problem (aka subgraph isomorphism problem)~\cite{reutter2013graph}. 
GPM can be done over both undirected and directed graphs respectively\footnote{In this work we will only focus on directed graphs.}. 
Traditionally GPM is a computational task that can be defined as the evaluation of graph patterns over a graph database~\cite{AnglesFMGQLs16}.
The most trivial form of a graph pattern is the basic graph pattern (BGP). 
A BGP coupled with features such as projection, union, difference and optional forms a complex graph pattern (CGP).
We illustrate these concepts in brief with respect to. context of SPARQL and Gremlin GQLs in Section~\ref{sec:gpm_gremlin}.
Detailed information on GPM is available in~\cite{AnglesFMGQLs16,ReinventingGubichevT14,reutter2013graph}.

GPM is carried out by \textit{matching} (also referred to as \textit{evaluation}), a sub-graph pattern over a graph $G$. 
Matching has been formally defined in various texts and we summarize a formal definition in our context which closely follows the definition provided by~\cite{ReinventingGubichevT14,AnglesFMGQLs16}.


\begin{definition}[Match of a Graph Pattern $\llbracket P \rrbracket_{G}$]\label{def:match}
A graph pattern $P = (V_{p}, E_{p}, \lambda_{p}, \mu_{p})$; is matching the graph G = $(V,E,\lambda,\mu)$, iff the following conditions are satisfied:
\begin{enumerate}[nosep]
     \item there exist mappings $\mu_{p}$ and $\lambda_{p}$ such that, all variables are mapped to constants, and all constants are mapped to themselves (i.e. $ \lambda_{p} \in \lambda, \mu_{p} \in \mu$),
     \item each edge \'e $\in E_{p}$ in $P$ is mapped to an edge $e \in E$ in $G$, each vertex \'v $\in V_{p}$ in $P$ is mapped to a vertex $v \in V$ in $G$, and
     \item the structure of $P$ is preserved in $G$ (i.e. $P$ is a sub-graph of $G$) \qed
\end{enumerate}
\end{definition}
The definition for \textit{matching} for edge-labeled graphs is analogous to that of the property graph (ref. Def.~\ref{def:match}): \textit{(i)} a mapping $m$ maps the constants to themselves and variables to constants; \textit{and} \textit{(ii)} the structure of $P$ is preserved in $G$ (example illustration ref. Figure~\ref{fig:property_graph}).

\subsection{SPARQL Query Language}\label{sec:sparql}
SPARQL is a declarative GQL which is a W3C recommendation and the query language of the RDF triple stores.
The building blocks of SPARQL are RDF triple patterns, consisting of subject, predicate, and object, where either of it can be a variable, literal value or IRI.
In this work, we do not consider the blank node semantics.

    \begin{definition}[SPARQL BGP]\label{def:bgp_sparql}
A SPARQL query defines a graph pattern to be matched against a given RDF graph. A \textit{basic graph pattern (BGP)} is a set of \textit{triple patterns}, $tp =(s^{'}, p^{'}, o^{'})$ where $s^{'} \in \{s,?s\},~ p^{'} \in \{p,?p\}$ and $o^{'} \in \{o,?o\}$.
\qed
    \end{definition}

\subsubsection{Evaluation of a graph pattern in SPARQL}
SPARQL operates over homomorphism-based bag semantics defined in~\cite{schmidt2010foundations,Angles16Powerset}.
In the context of SPARQL, the evaluation of a graph pattern $P$ against an RDF graph $G$ has been well defined in literature. 
We refer to \cite{prud2006sparql,perez2006semantics,angles2008expressive,Angles16Powerset,AnglesFMGQLs16} for the formal definitions.
In the later sections we illustrate the evaluation of a SPARQL graph pattern with examples.

\subsection{The Gremlin Graph Traversal Language and Machine}

Gremlin is the query language of \emph{Apache TinkerPop}\footnote{Gremlin: Apache TinkerPop's graph traversal language and machine~(\url{https://tinkerpop.apache.org/})} graph computing framework. 
Gremlin is system agnostic, and enables both -- pattern matching (declarative) and graph traversal (imperative) style of querying over graphs.

\subsubsection{The Machine.} Theoretically, a set of traversers in $T$ move (traverse) over a graph $G$ (property graph, cf. Section~\ref{def:pg}) according to the instruction in $\Psi$, and this computation is said to have completed when there are either: 
\begin{enumerate}
    \item no more existing traversers ($t$), or 
    \item no more existing instructions ($\psi$) that are referenced by the traversers (i.e. program has halted).
\end{enumerate}
The result of the computation is either an empty set (i.e. former case) or the multiset union of the graph locations (vertices, edges, labels, properties, etc.) of the halted traversers which they reference. Rodriguez~\cite{rodriguez2015traversalmachine} formally define the operation of a traverser $t$ as follows: 
\begin{equation}
    G \leftarrow \hspace{2pt} _{\mu}  \frac{t \in T}{\{\beta, \varsigma\}}  \hspace{2pt} _{\psi} \rightarrow \Psi{\text{~\cite{rodriguez2015traversalmachine}}}
\end{equation}
where, $\mu$: \textit{T} $\rightarrow$ \textit{U} is a mapping from the traverser to its location in G; $\psi$: \textit{T} $\rightarrow$ $\Psi$ maps a traverser to a step in $\Psi$; $\beta$: \textit{T} $\rightarrow$ $\mathbb{N}$ maps a traverser to its bulk\footnote{ The bulk of a traverser is number of equivalent traversers a particular traverser represents.}; $\varsigma$: \textit{T} $\rightarrow$ \textit{U} maps a traverser to its sack (i.e. local variable of a traverser) value. 

\subsubsection{The Traversal.}\label{sec:traversal}
A Gremlin graph traversal can be represented in any host language that supports function composition and function nesting. 
These steps are either of: 
\begin{enumerate}
    \item \textit{Linear motif} - $ f \circ g \circ h $, where the traversal is a linear chain of steps; or
    \item \textit{Nested motif} - $ f \circ ( g \circ h) $ where, the nested traversal $ g \circ h $ is passed as an argument to step \textit{f}~\cite{rodriguez2015traversalmachine}. 
\end{enumerate}
A step $ f \in \Psi $ can be, defined as $ f : A^{\star} \rightarrow B^{\star} $\footnote{The Kleene star notation ($A^{\star}, B^{\star}$) denotes that multiple traversers can be in the same element (A,B).}. 
Where, \textit{f} maps a set of traversers of type A (located at objects of A) to a set of traversers of type B (located at objects of B).
Given that Gremlin is a language and a virtual machine, it is possible to design another traversal language that compiles to the Gremlin traversal machine (analogous to how Scala compiles to the JVM).
As a result, there exist various Gremlin dialects such as Gremlin-Groovy, Gremlin-Python, etc. 

\subsubsection{Evaluation of a graph pattern in Gremlin}
In Gremlin, GPM is performed by traversing\footnote{The act of visiting of vertices ($v \in V$) and edges ($e \in E$) in a graph in an alternating manner (in some algorithmic fashion)~\cite{DBLP:books/igi/Sakr11/RodriguezN11}.} over a graph $G$. A traversal $t$ over $G$ derives paths of arbitrary length. 
Therefore, a GPM query in Gremlin can be perceived as a path traversal. 
Rodriguez et al.~\cite{DBLP:conf/icde/RodriguezN11} define a \textit{path} as:    

\begin{definition}[Path]\label{def:path}
    A path $p$ is a sequence or a string, where \textit{p} $\in E^{\star}$ and $E \subset (V \times L_{e} \times V$)\footnote{The kleene star operation $\star$ constructs the free monoid E$^{\star}$ = $\bigcup_{n=0}^{\infty}$ E$^{i}$. where E$^{0}$ = \{$\epsilon$\}; $\epsilon$ is the identity/empty element.}. A path allows for repeated edges and the length of a path is denoted by $||$\textit{p}$||$, which is equal to the number of edges in \textit{p}. \qed
\end{definition}

Moreover, from~\cite{DBLP:books/igi/Sakr11/RodriguezN11} we also know that these path queries are comprised of several atomic operations called the single-step traversals. We discuss these in brief in Section~\ref{sec:bgps_as_sst}. The evaluation of an input graph pattern in Gremlin is taken care by two functions: 
\begin{enumerate}
    \item the recursively defined \texttt{match()} function, which evaluates each constituting graph pattern and keeps track of the traverser's location in the graph (i.e. path history), and, 
    \item the \texttt{bind()} function, which maps the declared variables (elements and keys) to their respective values. 
\end{enumerate}
The evaluation (also know as matching, ref. Def.~\ref{def:match}) of a graph pattern in Gremlin is carried out by the \texttt{match()}-step. 
We borrow the notation of the \textit{evaluation} of a graph pattern ($\llbracket Q \rrbracket_{G}$) from~\cite{perez2006semantics} for representing the evaluation of a Gremlin traversal $\Psi$ over a graph $G$, i.e. $\llbracket\Psi\rrbracket^{gml}_{G}$.
Details of execution of the \texttt{match()-}step in Gremlin are described in \cite{rodriguez2015traversalmachine}.  
\section{SPARQL \texorpdfstring{$\leftrightarrow$}{} Gremlin homology}\label{sec:gpm_gremlin}
In this section we present the correspondence between SPARQL BGPs/CGPs with Gremlin pattern matching path traversals.
In doing so we devise a formal analogy borrowing the evaluation semantics of a SPARQL query~\cite{AnglesFMGQLs16,perez2006semantics,schmidt2010foundations} (referring to the well established bag semantics) and put them in context of Gremlin traversals~\cite{rodriguez2015traversalmachine,DBLP:books/igi/Sakr11/RodriguezN11,DBLP:conf/icde/RodriguezN11}.
A detailed discussion on the one-to-one operator level mapping between SPARQL and Gremlin can be found in the study~\cite{thakkar2017graph}.
Furthermore, we illustrate the applicability of these concepts with respect to the running examples (as shown in Figures~\ref{fig:rdf_graph} and~\ref{fig:property_graph}).

\subsection{Graph Pattern Matching via Traversing}\label{sec:gremlin_gpmq}
A SPARQL query consists of several BGPs which when used with features such as projection or union, form a CGP (as we discussed in section~\ref{sec:sparql}).
BGPs (ref. Definition~\ref{def:bgp_sparql}) are comprised of \textit{triple patterns}, which match to RDF triples that constitute the RDF dataset. 
Moreover, the RDF data model resembles essentially a directed, edge-labeled, multi-graph or RDF graph. 
It is, therefore, possible to traverse an RDF graph with Gremlin (i.e. construct traversals), regardless of it being an edge-labeled graph or a property graph as the core-concept of traversing remains unaltered.

Analogous to SPARQL, Gremlin also provides the GPM construct, using the \texttt{Match()}-step. 
This enables the user to represent the query using multiple individually connected or disconnected graph patterns. 
Each of these graph patterns can be perceived as a simple path traversal, to-and-from a specific source and destination, over the graph. 

In Gremlin, each traversal can be perceived as a path query, starting from a particular source (A) and terminating at a destination (B) by visiting vertices $v \in V$ and edges $e \in E (V \times V)$. 
Each path query is composed of one or more single-step traversals (SST) as shown by~\cite{rodriguez2015traversalmachine}. 
Through function composition and currying, it is possible to define a query of arbitrary length~\cite{rodriguez2015traversalmachine}.
Furthermore, just as multiple BGPs form a CGP in SPARQL, the corresponding SSTs can be coupled with features such as projection, union, optional, etc. to form a complex path traversal query in Gremlin.
These path queries can be a combination of either a source, destination, labeled traversal or all of them in a varying fashion, depending on the information need of the user.

\subsection{SPARQL BGPs as Gremlin Single Step Traversals}\label{sec:bgps_as_sst} 
In this section we establish the exact analogy between SPARQL BGPs and Gremlin SSTs.
In SPARQL, GPM is conducted by matching BGPs which consist of triple patterns (TP), that form a sub-graph, against an RDF graph $G$ (i.e. checking whether a sub-graph is contained in G).
We can represent BGPs notationally as:
\begin{equation}\label{eqn:BGP}
BGP = \texttt{\{$TP$ \}}^{+} \hspace{2pt} \text{;} \hspace{2pt}TP = \texttt{\{s p o . \}}^{*}\footnotemark
\end{equation}
\footnotetext{The $*$ symbolizes that a TP can also be an empty graph pattern, whereas $+$ symbolizes that a BGP can consists of more than one TPs (i.e. a set of triple patterns).}
In this unique representation, each subject (\texttt{s}) and object (\texttt{o}) (i.e. nodes) in a triple is connected through only one predicate (\texttt{p}) relation (i.e. edges).
Figure~\ref{fig:rdf_graph} presents an example of the graph representation of a sample RDF dataset.

In Gremlin, GPM is conducted by the \texttt{match()}-step, wherein each above graph patterns, represented as pattern matching traversals are evaluated against a graph G. 
We already know that Gremlin allows a user to form traversals of arbitrary length using function currying and composition.
Due to this functionality \textit{and} given the nature of the information represented in a triple, it is possible to represent the underlying traversal operation using a SST, which represented by its predicate/edge.

\begin{figure}[tb]
\centering
\includegraphics[width=0.55\textwidth]{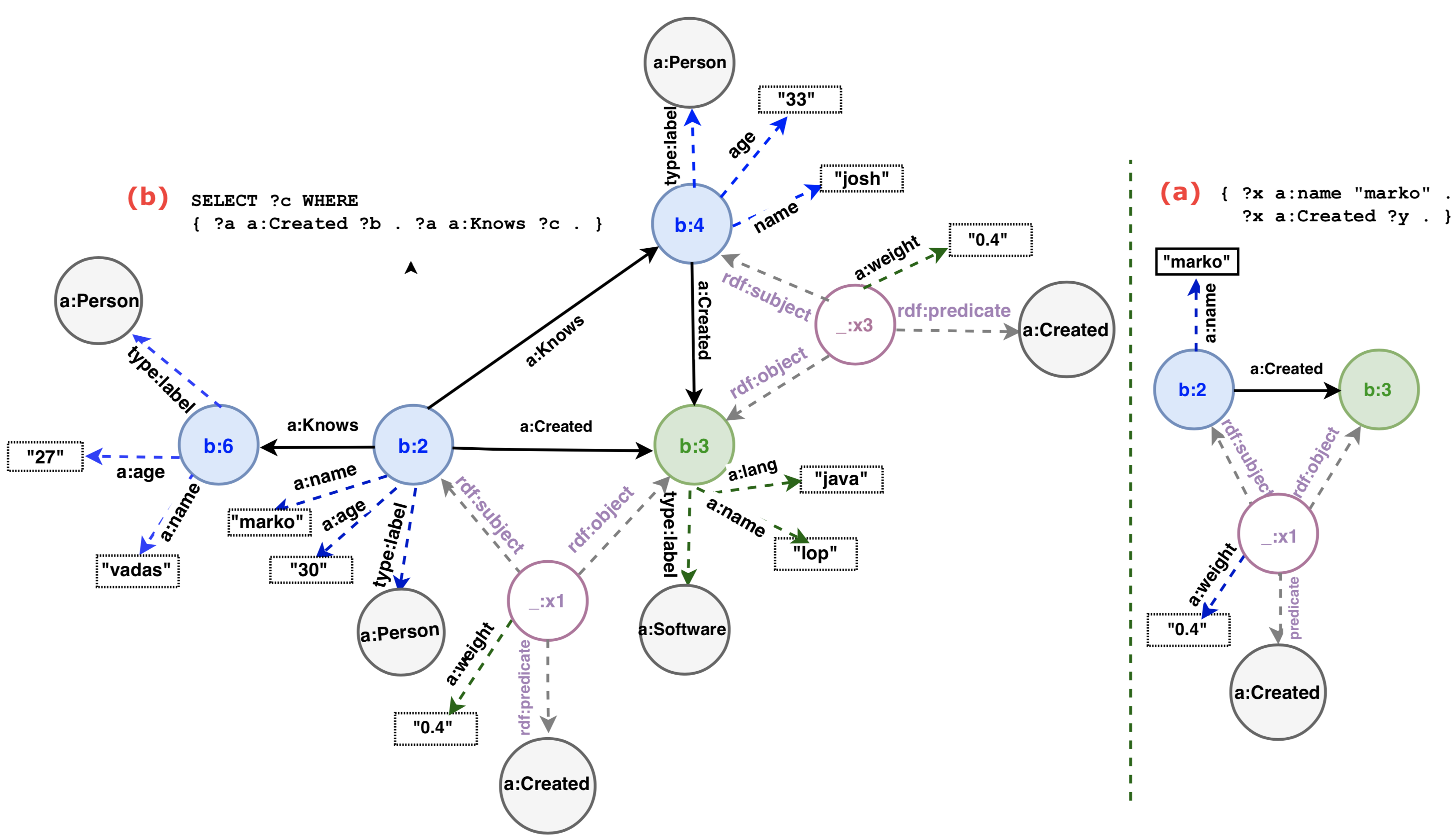} 
\caption{Example GPM evaluation notion of a, {\color{red}(a)} BGP and {\color{red}(b)} CGP, SPARQL query over an RDF graph in Figure~\ref{fig:rdf_graph}.}
\label{fig:example_1_rdf}
\end{figure}

For instance, consider the BGP in listing~\ref{lst1}, where the information need is to find what marko has created. This pattern, i.e. a subgraph formed by the BGP will be matched against a graph (ref. Figure~\ref{fig:rdf_graph}) to bind the values of the variables labeled as "x" to "marko", and "y" to the name property of the node connected by the edge/predicate labeled "created" by "x". 
Listing~\ref{lst1} represents the SPARQL BGP as shown in Figure~\ref{fig:example_1_rdf}{\color{red}(a)}.
\begin{lstlisting}[caption= {What has marko created?}, label={lst1}, language=Sparql]
{ ?x a:name "marko".  ?x a:Created ?y.}
\end{lstlisting}

The corresponding Gremlin traversal for the above SPARQL query is shown in listing~\ref{lst2} from Figure~\ref{fig:property_graph}.
Here the underlying SSTs required are \\ \texttt{.has('name','marko')} and \texttt{.out('Created')} that map to the \texttt{HasStep()} and \texttt{VertexStep()} instructions in the Gremlin instruction-set library~\cite{rodriguez2015traversalmachine,DBLP:conf/icde/RodriguezN11} respectively.
\begin{lstlisting}[caption= {An outgoing traversal from the vertex "marko" via an edge labeled "created".}, label={lst2}, language=Java]
g.V().as('x').has('name','marko').out('Created').as('y')
\end{lstlisting}
Here, \texttt{g.V()} i.e. $V_{g}$ is the traverser definition bijective to \textit{V} where, $\uplus_{i} \mu((V_{g})_{i})$ = V. 
Thus, each predicate in a triple pattern of a SPARQL BGP manifests the SST required for the matching the graph pattern.
We describe the different types of Gremlin SSTs and their correspondence with the SPARQL BGPs and summarize them in Table~\ref{tab:single step traversals}.

In~\cite{rodriguez2015traversalmachine}, Rodriguez presents an itemization of the Gremlin SSTs which can be combined together to form a complex path traversal (analogous to CGP in SPARQL).
We classify these SSTs into four categories depending on the whether the predicate-object combination \texttt{(s \underline{p o})} of the corresponding SPARQL BGP is a literal/value of a vertex/edge label (L) \textit{or} a vertex/edge property (P1) \textit{or} a variable representing a property value (P2) \textit{or} a traversal to and from a vertex given an edge label (E).
These four categories are:
\begin{itemize}[nosep]
\item \textbf{Case L} -- Traversal to access the label values of a vertex or an edge (L$_{v}$/L$_{e}$)
\item \textbf{Case P1/P2} -- Traversal to access the property values of a vertex \textit{or} an edge (P$_{v}$/P$_{e}$)
\item \textbf{Case E} -- Incoming/outgoing traversal between two adjacent vertices given an edge label (E$_{in}$/E$_{out}$)
\end{itemize}
We consider the above mentioned four cases as our base cases for constructing complex/composite traversals from SSTs with respect to their corresponding SPARQL CGPs.
Now, lets recall the SPARQL BGP from listing~\ref{lst1}. 
Here, the corresponding Gremlin SSTs for the SPARQL BGPs are the cases -- $P_{v}$ (as the traversal is to access the property value of a vertex) and $E_{out}$ (as the traversal is from a vertex named "marko" via the edge labelled "Created").
Table~\ref{tab:single step traversals} connects the dots by mapping the the Gremlin SSTs [notationally $\psi_{s}$] (defined in~\cite{rodriguez2015traversalmachine,DBLP:conf/icde/RodriguezN11}) to the SPARQL BGPs. 

\begin{table*}[tb]
\centering
\resizebox{0.95\textwidth}{!}{%
\begin{tabular}{llll}
\toprule
\textbf{{S.S.T.}~\cite{rodriguez2015traversalmachine}} & \textbf{Basic Graph Pattern (BGP)} & \textbf{Corresponding Gremlin Traversal Step} $\sigma${(}\textbf{BGP}{)} = $\psi_{s}$ & \textbf{Case} \\ 
\midrule
L$_{v}$ & \texttt{\{ ?x v:label "person" .\}} & {[}MatchStartStep(x), \textbf{HasStep({[}$\sim$label.eq(person){]})}, MatchEndStep{]}  &  \\
L$_{e}$ & \texttt{\{ ?x e:label "knows" .\}} & {[}MatchStartStep(x), \textbf{HasStep({[}$\sim$label.eq(knows){]})}, MatchEndStep{]}   & \multirow{-2}{*}{\textbf{L}}  \\ 
\hline
P$_{v}$ & \texttt{\{ ?x v:name "marko" .\}} & {[}MatchStartStep(x), \textbf{HasStep({[}name.eq(marko){]}}, MatchEndStep{]}   &  \\
P$_{e}$ & \texttt{\{ ?x e:weight "0.8" .\}} & {[}MatchStartStep(x), \textbf{HasStep({[}weight.eq(0.8){]})}, MatchEndStep{]} & \multirow{-2}{*}{\textbf{P1}} \\ 
\hline
P$_{e}$ & \texttt{\{ ?x e:weight ?y .\}} & {[}MatchStartStep(x), \textbf{PropertiesStep({[}name{]},value)}, MatchEndStep(y){]}   &  \\
P$_{v}$ & \texttt{\{ ?x v:name ?y .\}} & {[}MatchStartStep(x), \textbf{PropertiesStep({[}name{]},value)}, MatchEndStep(y){]}   & \multirow{-2}{*}{\textbf{P2}} \\ 
\hline
E$_{OUT}$ & \texttt{\{ ?x e:knows ?x .\}} & {[}MatchStartStep(x), \textbf{VertexStep(OUT,{[}knows{]},vertex)}, MatchEndStep(y){]}   &  \\
E$_{IN}$ & \texttt{\{ ?y e:knows ?x .\}}* & {[}MatchStartStep(y), \textbf{VertexStep(IN,{[}knows{]},vertex)}, MatchEndStep(x){]} & \multirow{-2}{*}{\textbf{E}} \\ \bottomrule
\end{tabular}
}
\caption{Mapping between the SPARQL BGPs, Gremlin SSTs and their corresponding Traversal steps. Each SPARQL BGP can be mapped to a corresponding Gremlin SST as described in Sect.~\ref{sec:bgps_as_sst}.}

\label{tab:single step traversals}
\end{table*}

\begin{figure}[tb]
\centering
\includegraphics[width=0.55\textwidth]{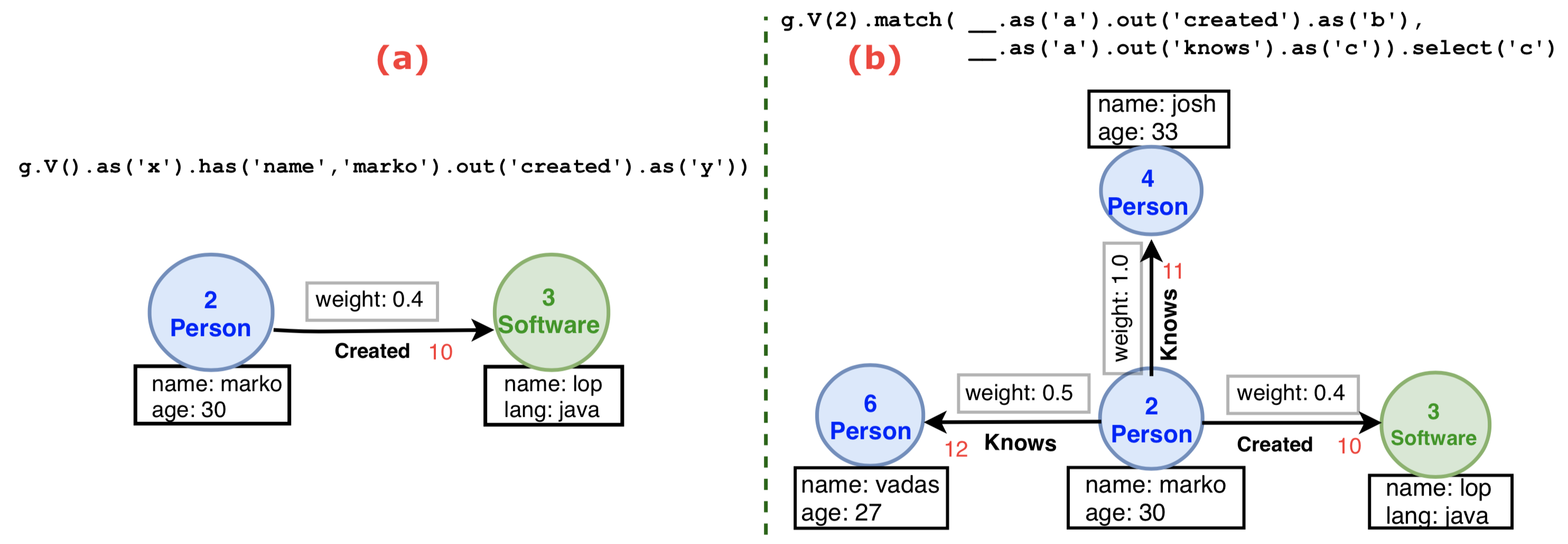}
\caption{Corresponding GPM evaluation notion of a, {\color{red}(a)} BGP and {\color{red}(b)} CGP, in a Gremlin traversal over a property graph in Figure~\ref{fig:property_graph}.}
\label{fig:example_1_pg}
\end{figure} 

\subsection{SPARQL Queries as Gremlin Pattern Matching Traversals}\label{sec:CGPs_as_path}

In SPARQL query language, as we have mentioned in earlier sections, a query comprises of one or more CGPs which in turn are formed by a combination of BGPs. 

Similarly, in Gremlin traversal language, a pattern matching traversal comprises of one or more path traversals which in turn are comprised of a combination of several SSTs.

From the already well established semantics of SPARQL query language~\cite{perez2006semantics,AnglesFMGQLs16,Angles16Powerset,cyganiak2005relational,harris2013sparql}, a query (Q) can be notationally represented as:
\begin{align}\label{eqn:CGP}
\begin{split}
Q = \{ \text{[PROJ.]} \hspace{2pt} BGP \hspace{2pt} \text{[UNION/DIFF./OPT.]} \\  BGP \hspace{2pt} \hspace{2pt} \text{[Filter (c)]}\}^{+}
\end{split}
\end{align}       
Where, Proj., Union, Diff. and Filter are the relational operators defined on the BGPs.

We have already established from Table~\ref{tab:single step traversals}, that each BGP can be mapped to a corresponding Gremlin single step traversal ($\sigma$(BGP) = $\psi_{s}$). Thus, from equations (\ref{eqn:BGP},~\ref{eqn:CGP}), we can create a mapping function $\sigma$, such that:
\begin{flalign}\label{eqn:bgp_psi}
\begin{split}
\sigma(BGP) = \psi_{s}     
\end{split}
\end{flalign}
Therefore, building on equations (\ref{eqn:bgp_psi},~\ref{eqn:CGP}) a SPARQL query $Q$ can be mapped as:
\begin{flalign}\label{eqn:query_psi}
\begin{split}
    \sigma(Q) &= \sigma\Big(\{ \text{\footnotesize[PROJ.]} \hspace{2pt} \textit{\footnotesize BGP} \hspace{2pt} \text{\footnotesize[UNION/DIFF./OPT.]} \hspace{2pt} \textit{\footnotesize BGP} \hspace{2pt} \text{\footnotesize[Filter (c)]}\}\Big)^{+}  \\
    &= \Big\{\sigma(\text{\footnotesize [PROJ.]}) \hspace{2pt} \psi_{s} \hspace{2pt} \sigma(\text{\footnotesize UNION/DIFF./OPT.}) \hspace{2pt} \psi_{s} \hspace{2pt} \sigma(\text{\footnotesize Filter (c)})\Big\}^{+} \\ 
    &= \Psi
\end{split}
\end{flalign}

Where, $\sigma(\text{[PROJ.]})$, $\sigma(\text{[UNION/DIFF./OPT.]})$ and $\sigma(\text{[FILTER (C)]})$ represent the respective Gremlin instruction steps for the operators such as Projection, Union, etc. 
We present a consolidated summary of the correspondence between SPARQL features/keywords and their corresponding Gremlin instruction steps in Table~\ref{tab:sparql_gremlin_keywords}. 
Furthermore, we also present the SPARQL query language constructs and their corresponding Gremlin traversal language constructs in Table~\ref{tab:sparql_gremlin_keywords}.

\begin{table*}[htbp]
\resizebox{\textwidth}{!}{%
\begin{tabular}{lllll}
\toprule
\textbf{Operation} & \textbf{SPARQL keyword} & \textbf{Gremlin Step} & \textbf{SPARQL construct (Q)} & \textbf{Gremlin construct $\sigma$(Q) = $\Psi$} \\ \midrule
Graph Pattern(s) & \{ \texttt{s p o . }\} & $\psi$ (i.e. \texttt{$\sigma$(s p o .)}) & BGP & $\psi$ (single step traversal {[}list of $\psi${]}) \\
Matching & \texttt{WHERE} \{ ... \} & \texttt{MatchStep(AND,{[}{]})} & WHERE \{ BGP1 . BGP2 . \} & {[}MatchStep(AND,{[}{[}$\psi_{1}${]},{[}$\psi_{2}${]}{]} \\
Restriction & \texttt{FILTER(C)} & \texttt{WhereTraversalStep(p(C))} & FILTER (?v1 \textless 30) & WhereTraversalStep({[}value(v1), IsStep(lt(30)){]}) \\
Join & \texttt{JOIN} & \texttt{AndStep()} & BGP1 * BGP2 & AndStep({[}{[}$\psi_{1}${]}, {[}$\psi_{2}${]}{]}) \\
Projection & \texttt{SELECT} & \texttt{SelectStep()} & SELECT ?v1 ?v2 & SelectStep({[}a, b,{]}) \\
Combination & \texttt{UNION} & \texttt{UnionStep()} & BGP1 UNION BGP2 & UnionStep(p(BGP1),p(BGP2)) \\
Deduplication & \texttt{DISTINCT} & \texttt{DedupStep()} & DISTINCT ?v1 & DedupStep({[}a,b{]}) \\
Restriction & \texttt{LIMIT(M)} & \texttt{RangeStep(0,M)} & LIMIT 2 & RangeStep(0,2) \\
Restriction & \texttt{OFFSET(N)} & \texttt{RangeStep(N,M+N)} & OFFSET 10 & RangeStep(10,12) \\
Sorting & \texttt{ORDER BY()} & \texttt{OrderStep()} & ORDER BY DESC(?a) & OrderStep({[}{[}value(a), desc{]}{]}) \\
Grouping & \texttt{GROUP BY()} & \texttt{GroupStep()} & GROUP BY(?a) & GroupStep(value(a)) \\  \bottomrule
\end{tabular}%
}
\caption{A consolidated list of SPARQL features/keywords \& their corresponding Instruction steps in Gremlin.}
\label{tab:sparql_gremlin_keywords}
\end{table*}

The evaluation of a SPARQL query $Q$ is carried out by matching or evaluating the graph patterns within Q, against a graph $G$ (an RDF graph in this case), denoted as $\llbracket P \rrbracket^{sparql}_{G})$.
Similarly, in Gremlin traversal language and machine, the evaluation of a pattern matching traversal $\Psi$ is carried out, by the \texttt{match()}-step, by matching or evaluating the SSTs within $\Psi$ against a graph $G$ (a property graph in this case). 
We borrow the same notation from~\cite{perez2006semantics,cyganiak2005relational} to fit our purpose and denote as $\llbracket\Psi\rrbracket^{gml}_{G})$.

We display brevity in constructing our arguments by quick examples instead of re-inventing the wheel by re-defining formal concepts and proofs (which already have been addressed in the works~\cite{Angles16Powerset,martonformalizing,holschNeo4jalgebra,AnglesFMGQLs16}). 
Moreover, we illustrate using examples, the semantic analogy between the evaluation of Gremlin traversal features in a homologous fashion to that of the multi-set semantics of SPARQL queries defined by~\cite{Angles16Powerset} who extend the work of~\cite{cyganiak2005relational,perez2006semantics}.
We show, by structural analogy created with the evaluation semantics of SPARQL\footnote{This is because both Gremlin and SPARQL operate over bag semantics and works such as~\cite{holschNeo4jalgebra,Angles16Powerset,schmidt2010foundations} have already debated and formally established the  equivalence between underlying semantics of relational and graph-specific operators for RDF and Property graphs}, that:
\begin{align}
\begin{split}
\llbracket\text{Q}\rrbracket_{G}^{sparql} \equiv \llbracket\Psi\rrbracket_{G}^{\tiny{gml}} \hspace{5pt} (\because \sigma(Q) = \Psi, \hspace{2pt} Eqn:~\ref{eqn:query_psi})
\end{split}
\end{align}

\textbf{Projection.}
The \textit{projection} operator projects/selects the values of a specific set of variables ($x,y,..,n$), from the solution of a matched graph pattern $P$, against the graph $G$.
Furthermore, it is possible to declare variables in Gremlin using \texttt{.as()} steps, which serve as syntactic sugars.
For instance, in the CGP as shown in Figures~\ref{fig:example_1_rdf}{\color{red}(b)} and~\ref{fig:example_1_pg}{\color{red}(b)}, we project only variable \texttt{?c} despite using (?a, ?b \& ?c) in the query, since we are only interested knowing the value binded to it. 
It is carried out using the \texttt{SELECT} keyword in SPARQL, and corresponding \texttt{.select()} step in Gremlin.
The corresponding evaluation of a select step in Gremlin can be illustrated as: 

\begin{flalign}\label{eqn:projection}
\begin{split}
&\RHD \text{$\llbracket$}\text{\color{blue}SELECT} \hspace{2pt} \texttt{?x1 ?x2} \hspace{2pt} \{ \textit{BGP} \} \rrbracket_{G}^{\tiny{sparql}} \\ &= \llbracket\sigma(\text{\color{blue}SELECT}  \texttt{?x1 ?x2} \hspace{2pt} \{ \textit{BGP} \})\rrbracket^{gml}_{G} \hspace{2pt} \\  &= \llbracket\sigma(BGP) \hspace{2pt} \sigma(\text{\color{blue}SELECT} \hspace{2pt} \texttt{?x1 ?x2})\rrbracket^{gml}_{G}   \\
&= \llbracket\psi_{s} \hspace{2pt} \texttt{\color{blue}SelectStep([x1,x2])}\rrbracket^{gml}_{G} =   \llbracket\Psi\rrbracket^{gml}_{G}
\end{split}
\end{flalign}
Here, $\psi_{s}$ (SST) and \texttt{SelectStep([x1,x2])} collectively form the final pattern matching traversal (analogous to a collection of BGPs and BGPs forming). 
Moreover, $\psi_{s}$ is mapped from Table~\ref{tab:single step traversals}, depending on the case it corresponds.

\textbf{Optional.}
The \textit{optional} operator in corresponds to a \textit{left-join} operation (in relational sense).
The optional graph patterns in a query are declared using this operator.
For instance, given the CGP: BGP1 \texttt{OPT.} BGP2; if the optional BGP2 does not match with graph G, then the results of BGP1 are returned unchanged, else additional bindings of BGP2 are added to the solution.
It is present in both SPARQL (as OPTIONAL) and Gremlin (as \texttt{.optional()} keyword which corresponds to \texttt{ChooseStep()} in the Gremlin instruction library).
\begin{align}\label{eqn:optional}
\begin{split}
&\RHD \llbracket BGP_{1} \hspace{2pt} \text{\color{blue}OPT.} \hspace{2pt} BGP_{2} \rrbracket_{G}^{\small{sparql}} \\ &=  \llbracket\sigma(BGP_{1} \hspace{2pt} \text{\color{blue}OPT.} \hspace{2pt} BGP_{2})\rrbracket^{gml}_{G}  \\ &= 
\llbracket\sigma(BGP_{1}) \hspace{2pt} \texttt{\color{blue}ChooseStep(} \hspace{2pt} \sigma(BGP_{2}) \text{\color{blue})}\rrbracket^{gml}_{G} \\
     &=  \llbracket \psi_{s1}, \texttt{\color{blue}ChooseStep(}\psi_{s2}\text{\color{blue})}\rrbracket_{G}^{gml} = \llbracket\Psi\rrbracket^{gml}_{G}
\end{split}
\end{align}

\textbf{Union.}
The \textit{union} operator combines the solution sets of the two input graph patterns. 
In SPARQL, union occurs between two BGPs or CGPs, analogously in Gremlin, it occurs between two SSTs and Traversals (i.e. the result set of two traversers). 
The solution set returned after the union operation is not de-duplicated by default, because of the governing bag semantics. 
Thus, all possible solutions are returned.
Formally, the evaluation of a union can be illustrated as:
 \begin{align}
 \begin{split}
      &\RHD \llbracket BGP_{1} \hspace{2pt} \text{\color{blue}UNION} \hspace{2pt} BGP_{2}\rrbracket_{G}^{\small{sparql}} \\ &=
      \llbracket\sigma(BGP_{1} \hspace{2pt} \text{\color{blue}UNION} \hspace{2pt} BGP_{2})\rrbracket_{G}^{\small{gml}} \\ &=  
      \llbracket \texttt{\color{blue}UnionStep(}\sigma(BGP_{1}) \text{,} \hspace{2pt} \sigma(BGP_{2}) \texttt{\color{blue}]}\rrbracket_{G}^{gml} \\
     &= \llbracket\texttt{\color{blue}UnionStep}([\psi_{s1}, \psi_{s2}])\rrbracket^{gml}_{G} = \llbracket\Psi\rrbracket^{gml}_{G}
 \end{split}
 \end{align}
For instance, consider the sample SPARQL CGP with UNION over the graph G (ref. Figure~\ref{fig:property_graph}) as illustrated in the example below.
The idea is to find the all the software created by "marko" which are in "java" language.
\begin{center}
\resizebox{0.48\textwidth}{!}{%
\begin{tabular}{ p{3.5cm} p{6.5cm} }
\multicolumn{2}{c}{\textbf{Illustration of a CGP with Union}} \\
\toprule
\textbf{SPARQL CGP} & \textbf{Gremlin Traversal ($\sigma$(BGP))} \\
\midrule
{\texttt{\{
?soft v:lang "java" .\} {\color{blue}{UNION}} \{ ?person v:name "marko" .\}}} & {\texttt{{\color{blue}UnionStep}
([[StartStep(soft), PropertiesStep([lang],value), IsStep(eq(java)),
EndStep], [StartStep@[person], PropertiesStep([name],value),
IsStep(eq(marko)), EndStep]])}}\\
\bottomrule
\end{tabular}
}
\end{center}

\textbf{FILTERs.}
The \textit{filter} keyword (or a group of operators) is used to restrict the results based on user-defined criteria.
Filters declare one or more constraints on the variables in the query, depending on the need of the user, and limit the solution of the overall group of BGPs with respect to  specified equality/inequality/regular expressions (i.e. constraints).
It is present in both SPARQL (as \texttt{FILTER C}, where \textit{C} is the declared constraint) and Gremlin (as \texttt{.where(C)}, where \textit{C} is the constraint).
In Gremlin the \texttt{.where(C)} keyword corresponds to the \texttt{WhereTraversalStep()} from the instruction set library

\begin{align}\label{eqn:filter-1}
\begin{split}
&\RHD \llbracket BGP \hspace{2pt} \text{\color{blue}FILTER} \hspace{2pt} C\rrbracket_{G}^{\small{sparql}} \\ &=  \llbracket\sigma(BGP \hspace{2pt} \text{\color{blue}FILTER} \hspace{2pt} C)\rrbracket_{G}^{gml} \\ &= \llbracket\sigma(BGP) \hspace{2pt} \sigma(\text{\color{blue}FILTER} \hspace{2pt} C)\rrbracket_{G}^{gml} \\
     &= \llbracket\psi_{s}, \hspace{2pt} \texttt{\color{blue}WhereTraversalStep}(\psi_{c})\rrbracket_{G}^{gml} \hspace{2pt} = \llbracket\Psi\rrbracket_{G}^{gml}
\end{split}
\end{align}
Here, $\psi_{c}$ denotes the corresponding Gremlin logical operator steps (i.e. \texttt{.eq()} for = , \texttt{.neq()} for $\neq$, \texttt{.gte()}, etc.). 
The Gremlin traversal language supports all the logical operators defined in SPARQL query language (as described here\footnote{SPARQL operator definitions --~(\url{https://www.w3.org/TR/rdf-sparql-query/\#SparqlOps})}), which can be found at the online documentation\footnote{Gremlin logical operators (predicates) --~(\url{http://tinkerpop.apache.org/docs/current/reference/\#a-note-on-predicates})}.
However, the current version of the Gremlin traversal language \textit{does not support} regular expression matching \texttt{REGEX} operators, although specific graph databases that leverage TinkerPop framework may provide a partial match extension. 

\begin{center}
\resizebox{0.48\textwidth}{!}{%
\begin{tabular}{ p{3cm}p{7cm} }
 \multicolumn{2}{c}{\textbf{Illustration of a CGP with FILTER}} \\
 \toprule
 \textbf{SPARQL CGP} & \textbf{Gremlin Traversal ($\sigma$(BGP))} \\
 \midrule{}
 {\texttt{\{ ?a v:name ?b . ?a v:age ?d . FILTER(?d<30) \}}} &
 {\texttt{[MatchStartStep(a), PropertiesStep([name],value),
 MatchEndStep(b)], [MatchStartStep(a),
 PropertiesStep([age],value)@[d],
 MatchEndStep(d)]}}, \texttt{{\color{blue}WhereTraversalStep}( [WhereStartStep(d),
 IsStep(lt(30))])} \\
 \bottomrule
\end{tabular}
}
\end{center}		    
Like in SPARQL, it is possible to declare multiple constraints inside a single FILTER clause: \\
FILTER (C1 {\color{blue}\&\&} C2) $\rightarrow$ \\ \texttt{WhereTraversalStep({\color{blue}AndStep}[(C1, C2)]} FILTER (C1 {\color{blue}$\mid\mid$} C2) $\rightarrow$ \\ \texttt{WhereTraversalStep({\color{blue}OrStep}[(C1, C2)]} \\
For brevity we skip the illustration of this step, as it being perceptible. 

\textbf{Query Modifiers.}
The solution set returned by the evaluated graph patterns is NOT \textit{de-duplicated} or \textit{ordered} by default, as both the languages operate on bag semantics. 
Therefore, \textit{query modifiers} or solution sequence modifiers are used for presenting the results in the desired order.
We list out query modifiers, their corresponding keywords and language constructs in Table~\ref{tab:sparql_gremlin_keywords}.
Examples of query modifiers include \textbf{DISTINCT} (for result de-duplication), \textbf{LIMIT} \& \textbf{OFFSET} (for restricting no. of results), \textbf{GROUP BY} (for grouping manipulation of result stream), \textbf{ORDER BY} (for ordering manipulation of result stream).

For brevity we skip the formal definitions of each modifier, rather illustrate their correspondence and applicability in Table~\ref{tab:sparql_gremlin_keywords}. 

\textbf{Subgraphs.}
Like in SPARQL query language, it is also possible to load/create and query \texttt{NAMED} graphs.
This can be achieved using the Gremlin \texttt{Subgraph()}-step.
It allows a user to create custom graphs based on specific graph patterns (vertices, edges and properties) and later query them using the same approach as described earlier in this section.

\section{Approach}\label{sec:approach}
In this section we discuss our proposed approach -- \textit{Gremlinator}, its execution pipeline and limitations in brief.
\subsection{Encoding SPARQL prefixes}
We encode the prefixes of SPARQL queries within Gremlinator implementation, in order to aid the SPARQL to Gremlin translation process. 
We define custom prefixes keeping in mind the four categories of SSTs (as stated in sec.~\ref{sec:bgps_as_sst}).
For instance, the standard \texttt{rdfs:label} prefix (which is generally a predicate) is represented as \texttt{e:label} or \texttt{v:label} (where e = edge and v =  vertex).
A similar procedure is followed for other three cases. 
    
\subsection{Gremlinator Architecture \& Algorithm}
\begin{figure}[tb]
\begin{center}
\includegraphics[width=\columnwidth]{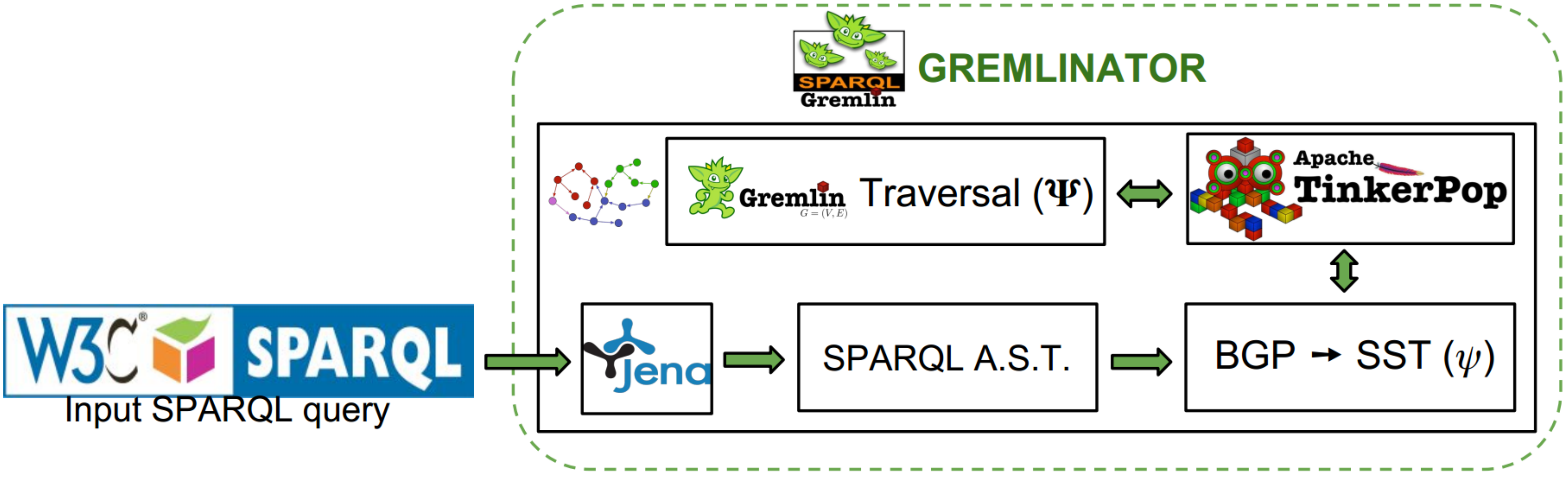}
\end{center} 
\caption{The architectural overview of Gremlinator.}
\label{fig:grem_arch}
\end{figure}

We now present the architectural overview of Gremlinator in Fig.~\ref{fig:grem_arch} and discuss the role of each of the four-step execution pipeline. 

\textbf{Step 1.} The input SPARQL query is first parsed using the Jena ARQ module, thereby: (i) validating the query and (ii) generating its abstract syntax tree (AST) representation. 

\textbf{Step 2.} From the obtained AST of the parsed SPARQL query, Gremlinator then visits each BGPs, mapping them to the corresponding Gremlin SSTs ($\psi_{s}$, ref. Table~\ref{tab:single step traversals}).

\textbf{Step 3.} Thereafter, depending on the operator precedence obtained from the AST of the parsed SPARQL query, each of the corresponding SPARQL keywords are mapped to their corresponding instruction steps from the Gremlin instruction library (ref. Table~\ref{tab:sparql_gremlin_keywords}). Thus, a final conjunctive Traversal ($\Psi$) is generated appending the SSTs and instruction steps. This can be perceived analogous to the SPARQL query language, wherein a set of BGPs form a single complex graph pattern (CGP).

\textbf{Step 4.} 
This final conjunctive traversal ($\Psi$) is used to generate bytecode\footnote{Bytecode is simply serialized representation of a traversal, i.e. a list of ordered instructions where an instruction is a string operator and a (flattened) array of arguments.} which can be used on multiple language and platform variants of the Apache TinkerPop Gremlin family.

\textbf{Algorithm.}
The SPARQL to Gremlin translation algorithm is presented in Algorithm~\ref{algo:1}
\begin{algorithm}[tb]
\DontPrintSemicolon
\SetAlgoLined
\SetKwInOut{Input}{input}\SetKwInOut{Output}{output}
\Input{$SQ:$ SPARQL Query}
\Output{$GT:$ Gremlin Traversal}
 $GT \leftarrow \emptyset$; T $\leftarrow \emptyset$\ \tcp*{list of single step traversals $T$}; 
 $AST \leftarrow \texttt{getAST(SQ)}$; BGPs $\leftarrow  \texttt{getAllBGPs(AST)}$\;
 \ForEach{$bgp_i \in BGPs$}{
 $T \leftarrow T \cup \psi_{s}$\ \tcp*{mapping BGP to Gremlin S.S.T. ($\psi_{s} = \sigma(bgp_i)$) $\because$ Table~\ref{tab:single step traversals}} 
 }
 \tcp*{mapping the corresponding Gremlin operators in the A.S.T.  cf. Table~\ref{tab:sparql_gremlin_keywords}}
 \If{$c \leftarrow AST.FILTER, \quad \exists c\neq \emptyset$}{ 
 \ForEach{$c \in AST $}{
 $T \leftarrow T \cup \texttt{WhereTraversalStep}(\psi_{c})$\; 
 } }
 \If{$c \leftarrow AST.UNION$}{ 
  $GT \leftarrow \texttt{UnionStep}(Match(T))$\;
   }
   \eIf{$|T|>1$}{
  $GT \leftarrow \texttt{Match(T)}$\;
   }{
   $GT \leftarrow GT \cup T$\;
 }
  \If{$c \leftarrow AST.ORDERBY$}{ 
  $GT \leftarrow T \cup \texttt{OrderStep}(\psi_{c})$\;
   }
  \If{$c \leftarrow AST.GROUPBY$}{ 
  $GT \leftarrow T \cup \texttt{GroupByStep}(\psi_{c})$\;
   }
   \If{$c \leftarrow AST.LIMIT$}{ 
   
    \eIf{$k \leftarrow AST.OFFSET$}{
        $GT \leftarrow T \cup \texttt{RangeStep}({k},{c}+{k})$\;
    }{
        $GT \leftarrow T \cup \texttt{RangeStep}({c})$\;
    }
   }
\Return{$GT$}\;
 \caption{SPARQL2Gremlin}
 \label{algo:1}
\end{algorithm}

\subsection{Limitations}
    The current version of Gremlinator supports the SPARQL 1.0 \texttt{SELECT} queries with the following exceptions: 
    1.) \texttt{REGEX} (regular expressions) in \texttt{FILTER} (restrictions) of a graph pattern are currently not supported\footnote{This is because the  \texttt{REGEX} feature is not supported in TinkerPop Gremlin as of now. Thus, it is Gremlin's limitation and not of our approach.}.
    2.) Gremlinator does not support variables for the property predicate, i.e.~the predicate \texttt{\{p\}} in a graph pattern \texttt{\{s p o .\}} has to be defined or known for the traversal to be generated. 
    This is because traversing a graph is not possible without knowing the precise traversal operation to the destination (vertex or edge) from the source (vertex or edge).
\section{Empirical Evaluation}\label{sec:evaluation}
We now shed light on the empirical evaluation settings of our experiments. 
These include the dataset and query descriptions, a carefully curated experimental setup (keeping in mind the various settings native to both RDF and Graph DMSs), a brief note on the correctness of our approach, the reported results and their meticulous discussion. 
Finally, with a brief note on the curated public demonstration of Gremlinator which promotes the users to get a first hand experience of the proposed system, we conclude the section.  

\subsection{Datasets}\label{sec:datasets}
\begin{figure}[ht]
\begin{center}
\includegraphics[width=0.5\textwidth]{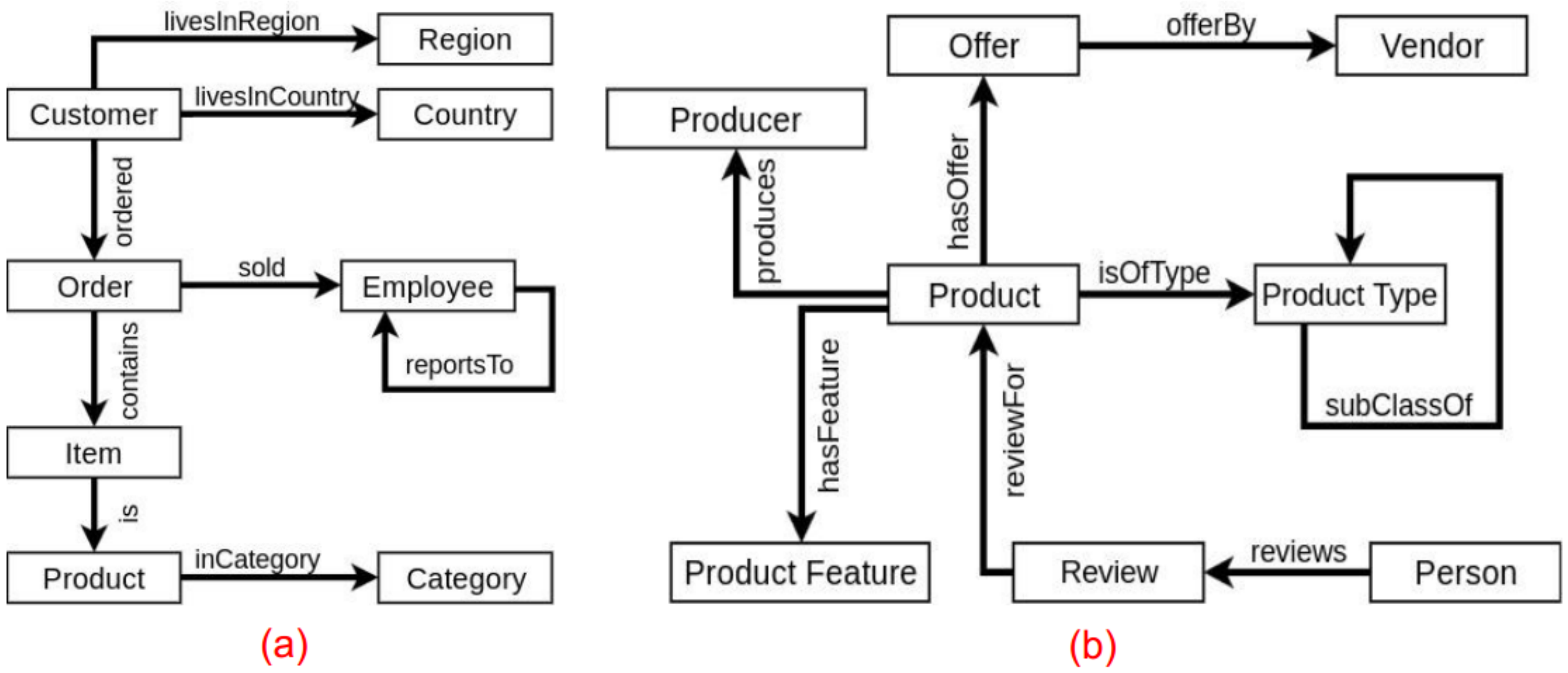}
\end{center} 
\caption{\textbf{The dataset schema of {\color{red}(a)} Northwind and {\color{red}(b)} BSBM.}}
\label{fig:northwind}
\end{figure}

\textbf{Northwind --} 
is a synthetic-dataset with an e-commerce scenario between a fictional company "Northwind  Traders", its Customers, and Suppliers.
Originated as a sample dataset shipped with Microsoft Access\footnote{Northwind Database~(\url{https://northwinddatabase.codeplex.com/})}, it raised to fame with an enormous demand for e-commerce use cases in benchmarking DMSs. 
In Figure \ref{fig:northwind}{\color{red}(a)} we present the dataset schema.
We obtained graph version of the dataset from here\footnote{SQL2Gremlin website -- (\url{http://www.sql2gremlin.com})}. \\
\textbf{Berlin SPARQL Benchmark~\cite{bsbm} (BSBM) -- }
is a synthetic dataset, which is built around an e-commerce use case, between a set of products, their vendors, consumers who review the products.
It is a widely famous for benchmarking RDF DMSs as it offers the flexibility of generating graphs of custom size and density.
We generated a standard 1M triples dataset using their data generation script, which makes it available in various formats (e.g. .nt, .csv, .sql, .ttl, etc)
Figure \ref{fig:northwind}{\color{red}(b)} describes the schema of the BSBM Dataset. 

Table~\ref{tab:both_stats} summarizes the statistics of both Northwind and BSBM-1M dataset.

\begin{table}[tb]
\centering
\caption{Dataset statistics} 
\label{tab:both_stats}
\resizebox{0.4\textwidth}{!}{%
\begin{tabular}{lllll}
\toprule
\textbf{Criterion} & \multicolumn{2}{l}{\textbf{Northwind}} & \multicolumn{2}{l}{\textbf{BSBM}} \\
& \textbf{RDF} & \textbf{PG} & \textbf{RDF} & \textbf{PG} \\
\midrule
Classes & 11 & - & 159 & -\\
Entities \& Nodes & 4413 & 3209 & 71015 & 92757\\
Distinct subjects & 4413 & -& 71017 & -\\
Distinct objects & 8187 & -& 166384 & -\\
Properties & 55 & 55 & 40 & 40 \\
Number of Triples \& Edges & 33003 & 6177  & 1000313 & 238309 \\
\bottomrule
\end{tabular}
}
\end{table} 




\subsection{Pre-defined Queries}
We created a pre-defined set of 30 SPARQL queries, for each dataset, which cover 10 different query features (i.e. three queries per feature with a combination of various modifiers).
These features were selected after a systematic study of SPARQL query semantics~\cite{AnglesFMGQLs16,perez2006semantics,schmidt2010foundations} and from BSBM~\cite{bsbm} explore use cases\footnote{BSBM Explore Use Cases~(\url{https://goo.gl/y1ObNN})} and Watdiv Query templates\footnote{Watdiv Query Features~(\url{http://dsg.uwaterloo.ca/watdiv/basic-testing.shtml})}.
A gold standard set of corresponding Gremlin traversals of the SPARQL queries was created by three Gremlin expert users, for a twofold validation of the traversals generated by our approach. 
We elaborate on the approach evaluation and correctness in the following sub-section.
Table~\ref{tab:query_characs}, summarizes the query design and the feature distribution within them that was used for our experiment.

\subsection{Experimental Correctness}\label{sec:exp_corr_val}

In order to validate the correctness of our approach empirically, we -- 
\textbf{(a)} loaded the RDF datasets in the three top of the line RDF DMSs and the corresponding Property graph datasets in the three top of the line Graph DMSs (cf. Section~\ref{sec:Exp_Setup}). 
Thereafter, we executed the SPARQL queries against the RDF DMSs and the corresponding Gremlinator translated Gremlin traversals against the Graph DMSs. 
We then compared the results returned by each of these queries for correctness; 
\textbf{(b)} compared the results returned by the Gremlinator translated traversals with respect to that returned by the hand crafted Gremlin traversals (gold standard queries curated by three Gremlin experts), for the corresponding SPARQL queries, for all the three Graph DMSs.

Having conducted the above validation, we observed that the results returned by the -- \textbf{(a)} RDF and Graph DMSs were equal. 
However, the representation of the returned results were distinct. 
The results returned by the SPARQL queries were in a tabular format, whereas those returned by the Graph DMSs were in a list of sets format.
We report this using a subset of results for BSBM dataset over both RDF and Graph DMSs in Table~\ref{tab:bsbm_1m_results} of Appendix~\ref{sec:app_a} for reference. 
Here, we can clearly observe that the results in both the cases are equal though having two different representations.
A complete set of all the results for both the datasets can be referred by vising the online resource described in the caption of Table~\ref{tab:bsbm_1m_results}.
\textbf{(b)} Gremlin translated traversals and the hand crafted Gremlin traversals were also equal. Thus, ensuring that the proposed SPARQL $\rightarrow$ Gremlin translation approach is correct, as it preserves the meaning of the original query (i.e. the information need of the input SPARQL is not manipulated in the translation process.)

\begin{table}[tb]
\centering
\caption{Query feature design and  description}
\label{tab:query_characs} 
\resizebox{0.5\textwidth}{!}{%
\begin{tabular}{llcccccc}
\toprule
\rot{\textbf{Query}} & \rot{\textbf{Feature}} & \rot{\textbf{FILTER}} & \rot{\textbf{COUNT}} & \rot{\textbf{LIMIT}} & \rot{\textbf{DISTINCT}} & \rot{\textbf{\# Patterns}} & \rot{\textbf{\# Proj. Vars.}} \\
\midrule
C1 & CGP &  & \checkmark  &  & \checkmark & 2 & 2\\
C2 & CGP &  &   &  & \checkmark & 1 & 1\\
C3 & CGP &  &   &  & \checkmark & 1 & 1 \\
F1 & CONDITION & \checkmark (1)  &   &  & & 3 & 3\\
F2 & CONDITION & \checkmark (2)  &   &  & & 3 & 3\\
F3 & CONDITION & \checkmark (1) &   &  & \checkmark & 2 & 1\\
L1 & RESTRICTION & \checkmark (1) &   & \checkmark & \checkmark & 4 & 2\\
L2 & RESTRICTION &  & \checkmark  &  & & 2 & 2\\
L3 & RESTRICTION &  &  \checkmark &  & & 2 & 2\\
G1 & GROUP BY &  & \checkmark  &  & \checkmark & 2 & 2\\
G2 & GROUP BY & \checkmark (1) &   &  & & 6 & 2\\
G3 & GROUP BY &  & \checkmark  &  & & 1 & 2\\
Gc1 & GROUP COUNT &  & \checkmark  & \checkmark & & 3 &2 \\
Gc2 & GROUP COUNT &  & \checkmark  &  &  & 2 & 2\\
Gc3 & GROUP COUNT &  & \checkmark  &  \checkmark & & 1 & 2\\
O1 & ORDER BY &  &   & \checkmark & & 1 & 1\\
O2 & ORDER BY & \checkmark (1) &   & & & 4 & 3\\
O3 & ORDER BY &  &   & \checkmark & \checkmark & 1 & 1\\
U1 & UNION & \checkmark (2) &   & \checkmark & & 8 & 1\\
U2 & UNION & \checkmark (2) &   & & & 6 & 2 \\
U3 & UNION & \checkmark (2) &   & & \checkmark & 4 & 1\\
Op1 & OPTIONAL & \checkmark (1) &   & & & 3 & 3\\
Op2 & OPTIONAL &  &   & \checkmark & \checkmark & 6 & 2 \\
Op3 & OPTIONAL & \checkmark (2) &   & & & 8 & 3 \\
M1 & MIXED &  & \checkmark & \checkmark & & 3 & 2 \\
M2 & MIXED &  & \checkmark & \checkmark & \checkmark & 2 & 2 \\
M3 & MIXED &  & \checkmark & \checkmark & & 4 & 2\\
S1 & STAR & \checkmark (1) &   & \checkmark & & 12 & 11\\
S2 & STAR & \checkmark (1) &   & \checkmark & & 5 & 4 \\
S3 & STAR & \checkmark (1) &   & & & 10 & 9 \\
\hline
\textbf{TOTAL} & \textbf{30 Q.} & \textbf{-} & \textbf{-} & \textbf{-} & \textbf{-} & \textbf{-} & \textbf{-} \\
\bottomrule
\end{tabular}
}
\end{table} 

\subsection{Experimental Setup}\label{sec:Exp_Setup}
We selected the following DMSs for the experiments:
\textbf{RDF DMS:} Openlink Virtuoso~\cite{virtuoso} [v7.2.4],  JenaTDB\footnote{Apache Jena TDB~(\url{https://jena.apache.org/documentation/tdb/index.html})} [v3.2.0],  4Store~\cite{harris20094store} [v1.1.5];
\textbf{Graph DMS:} TinkerGraph~\cite{home2016apache} [v3.2.3], Neo4J\footnote{Neo4J~(\url{https://neo4j.com/})} [v1.9.6], Sparksee\footnote{Sparksee -- formerly DEX~(\url{http://sparsity-technologies.com/\#sparksee})} [v5.1]. 
All the experiments were performed on a machine with the following configuration:
\textbf{CPU:} Intel$^{\text{\textregistered}}$ Xeon$^{\text{\textregistered}}$ CPU E5-2660 v3 (20 cores @2.60GHz), \textbf{RAM:} 128 GB DDR3, \textbf{HDD:} 512 GB SSD, \textbf{OS:} Linux 4.2-generic.

\textbf{Evaluation Metrics.}\label{sec:eva_metrics}
The following conditions and parameters were considered for reporting all  results.
\begin{itemize}
    \item Query execution time (in milliseconds or ms) considered is the average of 10 runs for each query (of both SPARQL and translated Gremlin traversals).
    \item Queries executed in both cold and warm cache settings for respective DMSs. Where a \emph{warm cache}: implies that the cache is not cleared after each query run, and \emph{cold cache}: implies that the cache is cleared using the \texttt{'echo 3 > /proc/sys/vm/drop\_caches'} UNIX command after each query execution.
    \item For Graph DMSs, query execution time is recorded for both \textit{with} and \textit{without} creating explicit indices. We elaborate on the reason for the same, next.
\end{itemize}

\textbf{Indexing in RDF Triple Stores vs Graph DMS.} 
RDF triple stores typically index data employing pre-defined indices.
However, it is theoretically possible to have an RDF DMS totally index-free, but this would imply performing a linear search through the entire dataset (set of triples) for each query that is executed.
For this reason, having some pre-defined index setting within a RDF DMS by default is salient.
The same, however, cannot be said for Graph DMS wherein these indices have to be created manually, depending upon the use case. 
For instance, Openlink Virtuoso maintains two all-purpose full (bitmap indices over PSOG, POGS) and three partial indices (over SP, OP GS) in the default configuration\footnote{RDF indexing scheme in Virtuoso~(\url{http://docs.openlinksw.com/virtuoso/rdfperfrdfscheme/})}. Furthermore, 4Store in its default setting maintains a set of three full indices (R, P, M)~\cite{harris20094store}, where -- the \emph{R-index} is a hash-map index over RDF resources (URIs, Literals, and Blank Nodes); the \emph{P-index} consists of a set of two radix trees per predicate, using a 4-bit radix; the \emph{M-index} is a hash-map based indexing scheme over RDF Graphs (G). 
Lastly, Apache Jena TDB maintains three indices using a custom persistent implementation of B+ Trees\footnote{RDF indexing scheme in Apache Jena TDB~(\url{https://jena.apache.org/documentation/tdb/architecture.html\#triple-and-quad-indexes})}. 

On the other hand, Graph DMSs rarely maintain any default indexing scheme. 
They rather offer the possibility of creating explicit indexes over custom graph elements, using a variety of data structures, depending on the implementation. 
For instance, TinkerGraph supports the creation of regular and composite hash-map indices (multiple key-value pairs) on graph elements (node and edge attributes). 
Neo4J allows declaring regular indices (composite indices are supported from v3.5 onwards) on graph elements (including labels). 
It offers a variety of indices ranging from Lucene index (for textual attributes) and as SBTREE-based index (numeric ones, such as IDs), which is based on custom implementation of B-Trees with several optimizations related to data insertion and range queries~\footnote{Indexing in Neo4J~(\url{http://neo4j.com/docs/developer-manual/current/cypher/schema/index/})}. Lastly, like other Graph DMSs, Sparksee also offers user-defined indices on attributes. It uses a bitmap index implemented using sorted B-trees~\cite{martinez2011dex}.

As we pointed out earlier, it is not possible to have a completely index-free RDF DMS. 
Thus, in order to grasp a better understanding of query execution performance with respect to various factors (such as indexing schemes, query typology and cache configuration) and also for the sake of fairness (towards Graph DMSs) we run all the experiments with two settings of Graph DMSs, i.e. with (i.e. manually created) and without indices. 

\subsection{Results}\label{sec:results_discussion}

The detailed results including the queries, dataset statistics, plots and full configuration settings can be obtained from here\footnote{Detailed  results can be found at~(\url{https://goo.gl/CSSVzZ})}. 
The complete source code of Gremlinator is made publicly available, along with a recorded demonstration of Gremlinator in action, which can be accessed here\footnote{Gremlinator source code~(\url{https://github.com/LITMUS-Benchmark-Suite/sparql-to-gremlin})}.
The complete setup including all the datasets, scripts, and DMSs can be found here\footnote{Experimental setup~(\url{https://github.com/harsh9t/SWJ-2018-Experiments})}.
The average time for translating a SPARQL query to Gremlin traversal is 14 ms for BSBM and 12.5 ms for Northwind queries respectively.
\begin{figure*}
  \centering
  \includegraphics[width=\textwidth]{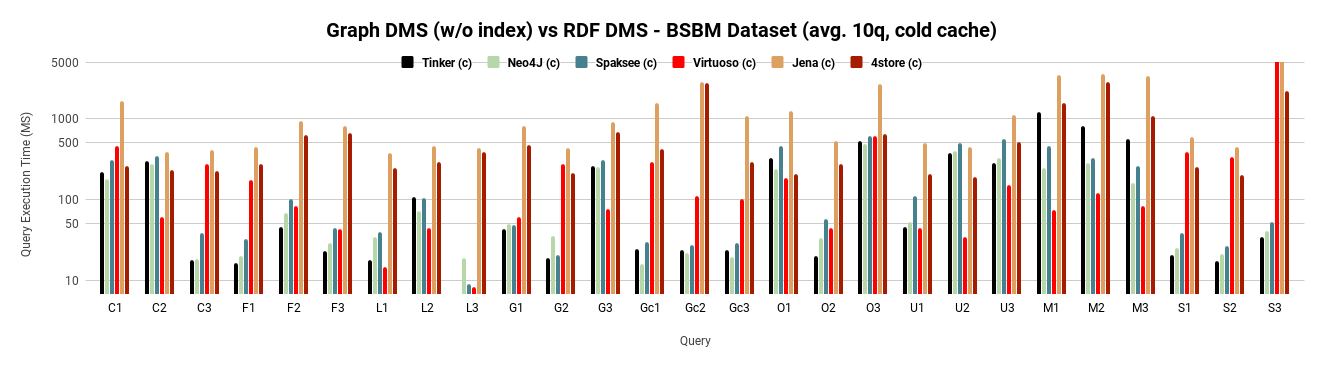}

  \includegraphics[width=\textwidth]{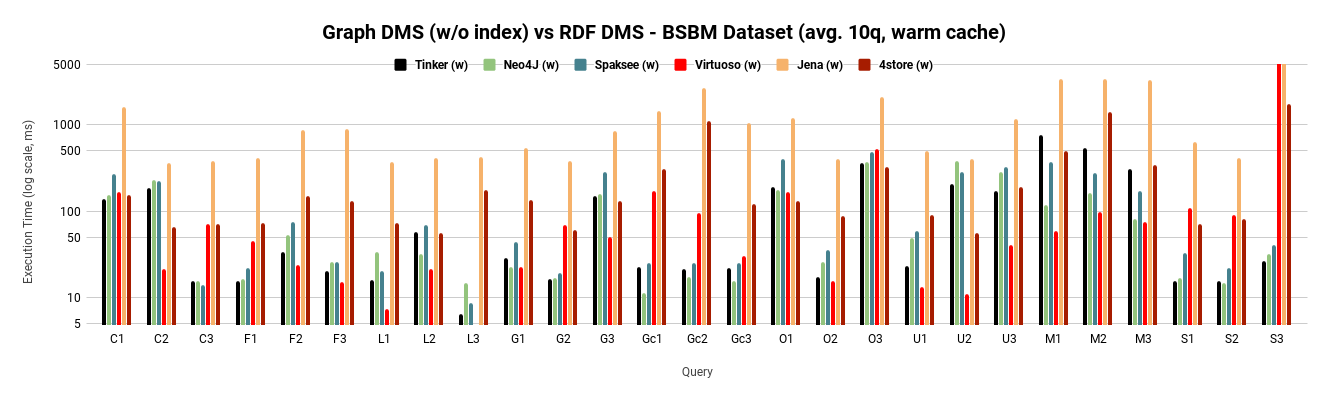}

  \includegraphics[width=\textwidth]{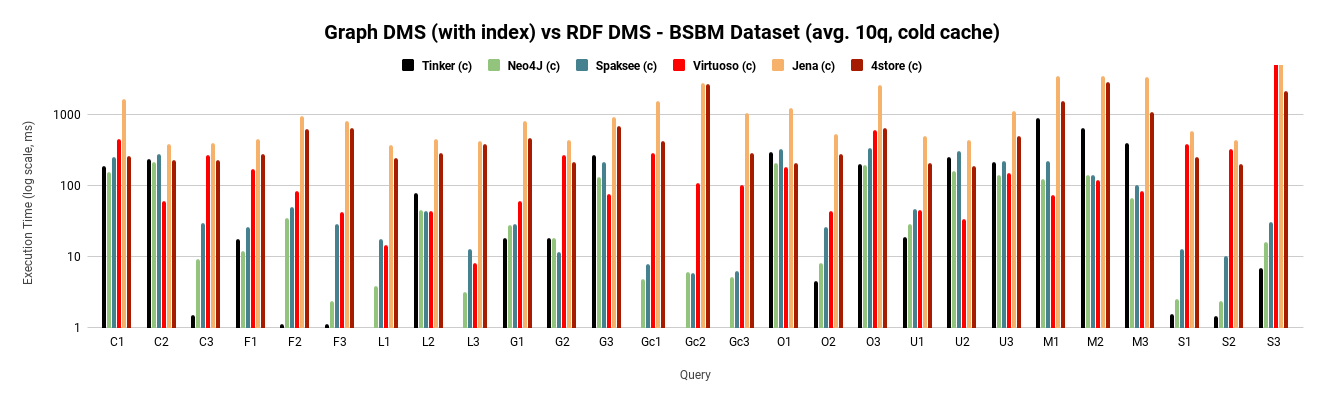}

  \includegraphics[width=\textwidth]{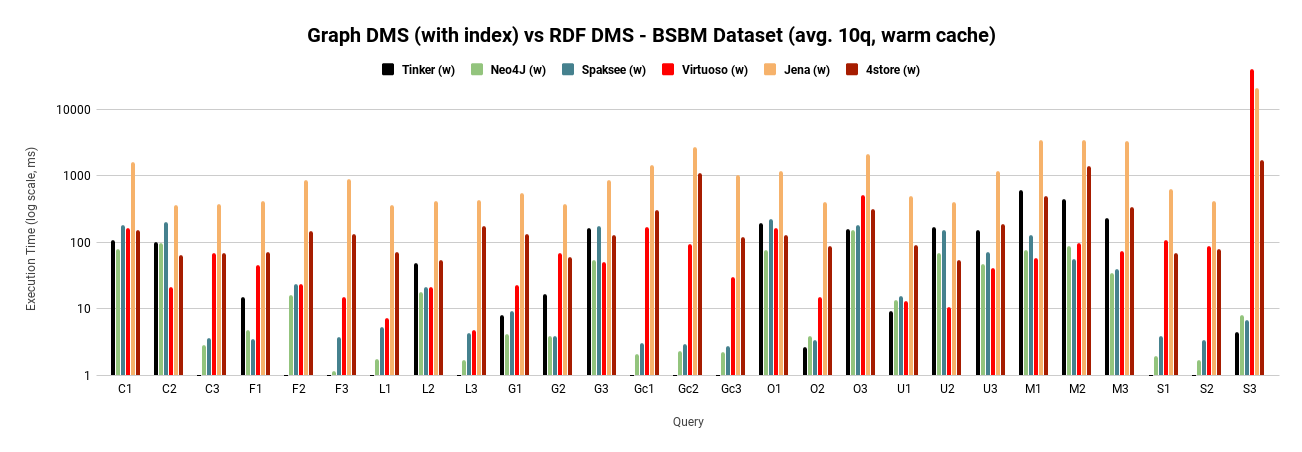}
  
   \caption{\textbf{Performance comparison of SPARQL queries vs Gremlin (pattern matching) traversals for BSBM dataset with respect to RDF and Graph DMSs in different configuration settings.}}
   \label{fig:bsbm_all_results}
\end{figure*}

\begin{figure*}
  \centering
  \includegraphics[width=\textwidth]{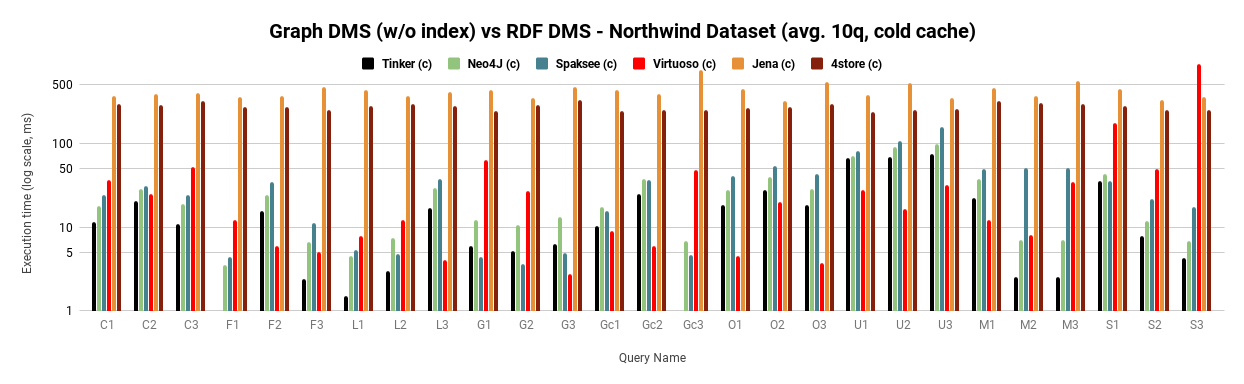}

  \includegraphics[width=\textwidth]{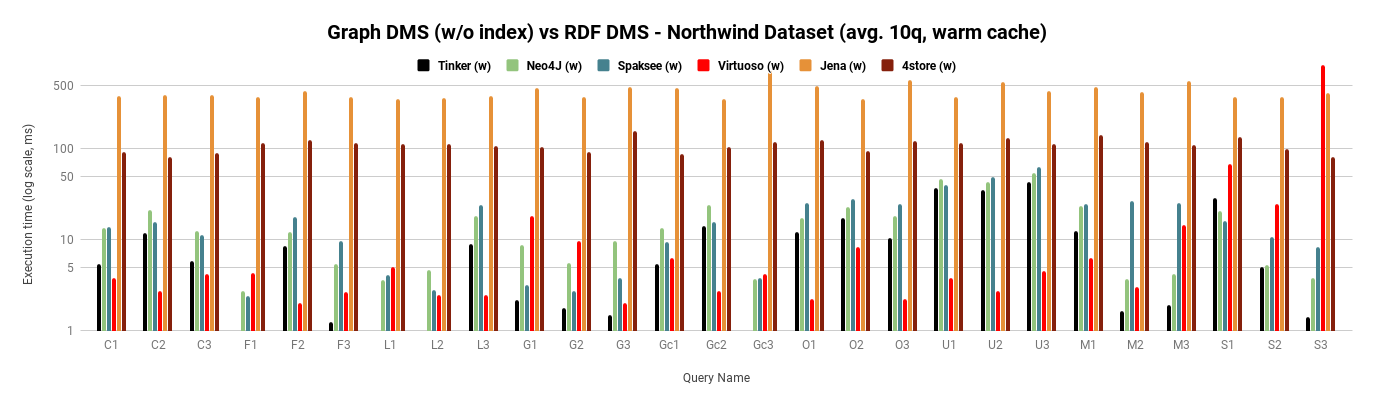}

  \includegraphics[width=\textwidth]{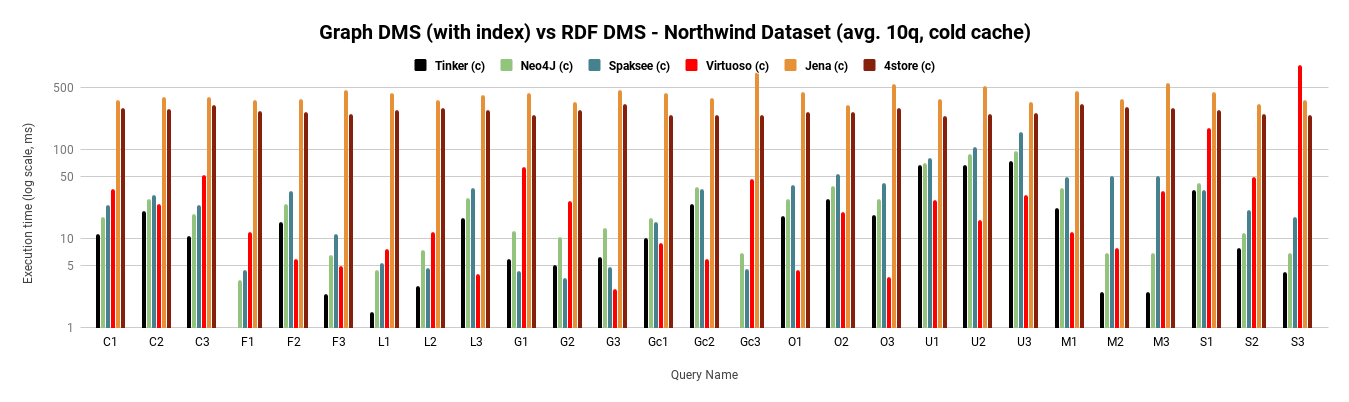}

  \includegraphics[width=\textwidth]{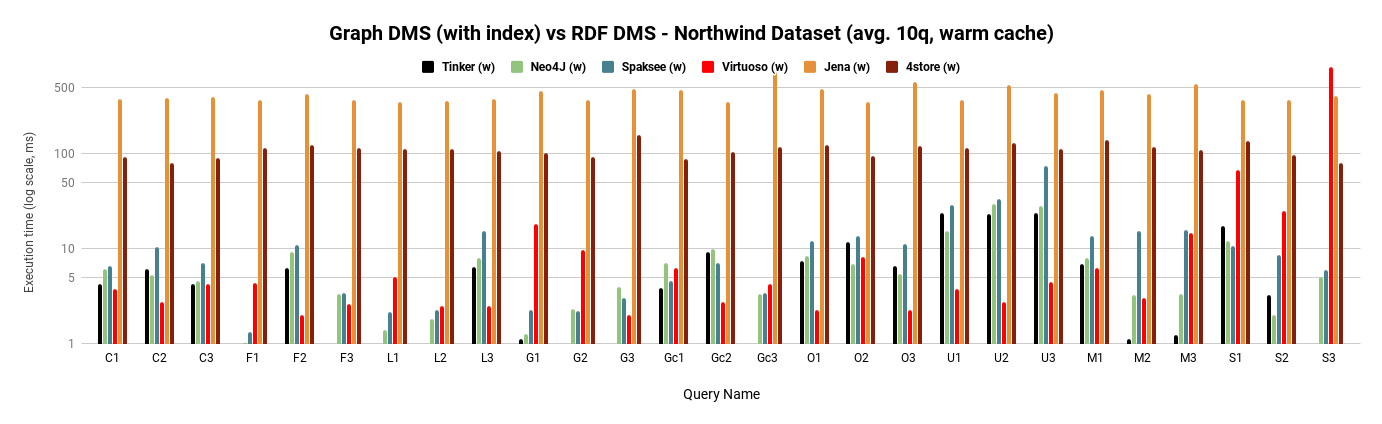}
 
   \caption{\textbf{Performance comparison of SPARQL queries vs Gremlin (pattern matching) traversals for Northwind dataset with respect to RDF and Graph DMSs in different configuration settings.}}
   \label{fig:nw_all_results}
\end{figure*}

Figures~\ref{fig:bsbm_all_results} and~\ref{fig:nw_all_results}, presents the plots of our experimental results, in all four settings, for the BSBM and Northwind datasets respectively. 
The plots follow log scale for execution time (in ms).
Furthermore, we also report the detailed query-wise results in tabular format in  Appendix~\ref{sec:app_det_res}, for more comprehensive understanding.
We observe similar trend of performance of SPARQL vs. Gremlin queries over both the datasets, which is evident from Figures~\ref{fig:bsbm_all_results} and~\ref{fig:nw_all_results} and also the Tables~\ref{tab:bsbm_full_stats} and~\ref{tab:nw_full_stats} of Appendix~\ref{sec:app_det_res}.
Therefore, we present the detailed performance analysis of SPARQL vs. Gremlin only for the BSBM dataset.
We organize our observations on the performances of participating DMSs as follows, and present our discussion.
    
    \textbf{Graph DMSs without index:}\label{sec:without_index}
    We categorize our findings in two groups -- cold cache and warm cache.
    We observe that for --
    \begin{enumerate}
        \item \textbf{Cold cache}: SPARQL queries report a comparative advantage with respect to Gremlin traversals, leveraging the advantage of indexing schemes of RDF DMSs. SPARQL performs moderately faster (1x-2x) for simple queries (C1, C2) and order by (O1, O3); substantially faster (3x-5x) for union and mixed queries (U1-3, M1-3).
        Whereas, Gremlin traversals benefit from \textit{only} the graph locality inherent to Graph DMSs. Gremlin traversals perform moderately faster (1x-2x) for restriction (L1, L3), group by (G1-3) and conditional (F1-3) queries;  substantially faster (3x-5x) for group count (Gc1-3) and star (S1-3) queries. 
        We also note that aggregation queries (counts, group counts) in Graph DMSs are an order of magnitude faster as compared to RDF DMSs since they do not have to execute multiple inner joins in addition to the aggregation operations.
        Moreover, for star-shaped queries (queries with bushy plans having >=5 triple patterns, >=1 filter and >=4 projection variables) Gremlin pattern matching traversals outperform their SPARQL counterparts by at least an order of one magnitude for S1, S2 and at least an order of two magnitudes for S3 (with 10 triple patterns, 1 filter and 9 projection variables).
        \item \textbf{Warm cache}: SPARQL queries reap the most benefits of warm caching from RDF DMSs as compared to the Gremlin traversals from Graph DMSs.
        We observe that on average, in this setting, the improvement is up to 1x-1.8x for star and mixed queries, 2x-3x for aggregation (counts), condition (filter) and re-ordering (order by, group by) queries, and 3x-5x for CGPs and union queries.
        We also note that SPARQL queries are almost an order of magnitude faster than the corresponding Gremlin traversals for queries having a union operator, and are comparable for mixed, CGPs, and order by queries. 
        Furthermore, we also note that SPARQL star-shape based queries do not register substantial improvement in warm cache execution.
        On the other hand, Gremlin traversals receive little benefit, from Graph DMSs, in warm cache. 
        We report that on average, in this setting, the improvement is up to 1.3x for aggregation (count, group count) and star-shaped queries; up to 1.5x for re-ordering (order-by, group-by) and condition (filters) queries; up to 2x for mixed, union and restriction (limit) queries. 
    \end{enumerate}
    
    \textbf{Graph DMSs with indexing:}
    We manually created composite indices for each Graph DMS on attributes such as \texttt{"name"}, \texttt{"customerId"}, \texttt{"unitPrice"}, \texttt{"unitsInStock"}, \texttt{"unitsOnOrder"} for BSBM dataset. Similarly, on \texttt{"type"}, \texttt{"productID"}, \\ \texttt{"reviewerID"}, \texttt{"productTypeID"} for Northwind dataset, on the node attributes (numeric) \footnote{We have provided all the groovy scripts used for creating composite indices in the Github repository pointed earlier}. The indices use the hash-map data structure. We did not re-execute SPARQL queries on RDF DMSs, as there was no change in the indexing setting for the same.
    \begin{enumerate}
    \item \textbf{Cold cache:} Gremlin traversals perform significantly faster when executed on Graph DMSs with composite indices. 
    We observe that, as compared to the previous (cold cache + without index) setting, the improvement reported on an average is up to 1x-2x for union, mixed and group by traversals; up to 2x-3x for re-ordering (group-by, order-by) traversals; up to 3x-5x for regular and restriction traversals; and >5x for aggregation and star-shaped traversals.
    \item \textbf{Warm cache:} In this setting the Graph DMSs (i.e. Gremlin traversals) register similar performance gains to that in non-indexed configuration.
    \end{enumerate}
    

\subsection{Discussion.}\label{sec:discussion}
We now discuss the findings of our experiments with respect to the factors which influence the query execution performance of a particular DMS and summarize our observations.
We categorize our findings based on the following factors:
\begin{itemize}[nosep]
    \item \textbf{Query typology:} 
    We report that for -- (i) simple/linear queries (such as C1-3, F1-3, L1-3) both SPARQL and Gremlin traversal performances are comparable; 
    (ii) SPARQL outperforms corresponding Gremlin traversals for union queries. 
    This is so because in SPARQL a union occurs between two or more sets of triple patterns.
    Whereas in the declarative construct (pattern matching) of Gremlin, a union occurs between two \texttt{.match()}-steps (i.e. Gremlin treats each \texttt{.match()}-step as a distinct traversal and then executes a union on top of it);
    (iii) Whereas, for complex queries (such as star-shaped and aggregation based queries), Gremlin traversals outperform their SPARQL counterparts. 
    As mentioned before (ref.~\ref{sec:without_index} -- cold cache section), this is because Graph DMSs do  not have to perform expensive joins (like RDF DMSs) on top of executing aggregation operations.
    (iv) Lastly, we also observe that for queries with greater number of projection variables (Proj. vars >= 3) and query modifiers (count, distinct, limit + offset, filter), Gremlin traversals show a distinctive advantage (more than an order of magnitude) in terms of performance with respect to corresponding SPARQL queries (e.g. for F1, F2, O2, S1, S2, S3). 
    This advantage, while still exists, is not as pronounced when comparing queries with a fewer number of projection variables and query modifiers.

    \item \textbf{Query caching -- Cold vs Warm:}
    Despite the fact that both DMSs benefit from warm cache query execution (as compared to cold cache), SPARQL queries receive the most advantage as compared to corresponding Gremlin traversals. 
    One reason for this is that Gremlin traversals perform considerately better (except in cases of union queries) by leveraging the advantage of underlying property graph data model (locality) and cannot be optimized further without explicitly creating regular or composite indices. 
    Out of all the three RDF DMSs, Jena shows the most gain in warm execution time, which receives up to 5x boost in cases such as union and CGP queries.
    
    \item \textbf{Indexing scheme:}
    It does not go without noticing, the one-sided dominance of Openlink Virtuoso, amongst all the evaluated RDF DMSs.
    As mentioned earlier, Virtuoso maintains a variety of full and partial indices. 
    Moreover, we also know that virtuoso employs custom partition clustering and caching schemes on top of these indices to provide an adaptable solution to all kinds of workloads.
    One distinctive advantage in virtuoso is that the indices are column-wise by default\footnote{Indexing scheme in Openlink Virtuoso~(\url{http://docs.openlinksw.com/virtuoso/rdfperfrdfscheme/})}, which takes one-third amount of space as compared to the row-wise indices.
    On the contrary, similar claims cannot be made about other RDF DMSs such as 4Store and JenaTDB.
    Graph DMSs, have a limited number options in terms of underlying indexing data structures implementation for creating manual indexes in the chosen version. 
    One reason can be deduced that there has not been an explicit need for using complex index schemes (as in Virtuoso), since composite indices based on B+ trees and hash-maps provide sufficient performance boost for graph traversal operations.
\end{itemize}

Thus, based on our findings, we  summarize that for complex queries (such as aggregation, star-shaped, and queries with higher number of projection variables + query modifiers) corresponding Gremlin pattern matching traversals outperform SPARQL queries. Whereas, for union-based queries SPARQL register significant performance advantage.

\subsection{Hands-On Gremlinator}
For the demonstration of our approach -- \textit{Gremlinator}, we provide the entire setup including both the datasets and the entire set of pre-defined SPARQL queries for interested users to get a first hand experience~\cite{ThakkarGremlinator2018}.
Furthermore, we encourage the end user to write and execute custom SPARQL queries for both the datasets, for further exploration.
As a part of the demonstration of our system~\cite{ThakkarGremlinator2018}, we provide-- 
    \textbf{(i)} an online screencast\footnote{Gremlinator Demo Tutorial --~\url{https://youtu.be/Z0ETx2IBamw}} for an introductory video tutorial on how to use the demonstration
    \textbf{(ii)} a web application accessible at\footnote{Gremlinator Web Demo --~\url{http://gremlinator.iai.uni-bonn.de:8080/Demo}}
    \textbf{(iii)} a desktop application of Gremlinator (standalone .jar bundle) which requires Java 1.8 JRE installed on the corresponding host machine, downloadable from the web demo website.
\section{Conclusion and Future Work}\label{sec:conclusion}
In this paper, we presented Gremlinator, a novel approach for supporting the execution of SPARQL queries on property graphs using Gremlin pattern matching traversals.
Furthermore, we presented an empirical evaluation of our approach using state-of-the-art RDF and Graph DMSs, demonstrating the validity and applicability of our approach.
The evaluation demonstrates the substantial performance gain obtained by translating SPARQL queries to Gremlin traversals, especially for star-shaped and complex queries. 
Gremlinator has obtained clearance by the Apache Tinkerpop development team and is currently in \href{https://github.com/apache/tinkerpop/tree/TINKERPOP-1878}{production phase} to be released as a plugin during TinkerPop's next framework cycle. 
Gremlinator has also been integrated into the \href{http://sansa-stack.net/}{SANSA Stack}~\cite{iswc_sansa} (v0.3) framework as an experimental plugin. 
Furthermore, Gremlinator is freely available under the Apache 2.0 license for public use from the \href{https://mvnrepository.com/artifact/io.github.litmus-benchmark-suite/gremlinator}{Maven Central} repository. 

As future work, we aim to -- 
(i) extend our current work by enabling support for SPARQL 1.1 featureset, such as Property Paths,  \textit{regex} in restrictions (i.e. FILTERs) and variables for property predicates; 
(ii) integrate Gremlinator within frameworks such as LITMUS~\cite{DBLP:LITMUS,thakkar2017trying,keswanilitmus}, to enable automatic execution of SPARQL queries over property graphs for robust benchmarking diverse RDF and Graph DMSs.

\begin{acks}
\vspace{-10pt}
	This work is supported by the funding received from EU-H2020 \href{http://wdaqua.eu/}{WDAqua ITN} (GA. 642795). We would like to thank Dr. Marko Rodriguez, Mr. Stephen Mallette, and Mr. Daniel Kuppitz, of the Apache TinkerPop project, for their support and quality insights for integrating Gremlinator.
\end{acks}

\begin{appendix}
\section{{\color{cyan}SPARQL - Gremlin Results}}\label{sec:app_a}
In this section we demonstrate the correctness of Gremlinator empirically, as already discussed in Section~\ref{sec:exp_corr_val}. 
We present a subset of the results in Table~\ref{tab:bsbm_1m_results}, which validate our claim that the proposed SPARQL $\rightarrow$ Gremlin translation is correct.
\begin{table*}[ht]
\centering
\resizebox{\textwidth}{!}{%
\begin{tabular}{|p{0.4cm}|p{7cm}|p{0.9cm}|p{7cm}|p{7cm}|}
\toprule
\textbf{Q.\#} & \textbf{SPARQL Query} & \textbf{Feature} & \textbf{SPARQL Query Result} & \textbf{Gremlin Traversal Result} \\
\midrule
C1 & \small{SELECT (COUNT (DISTINCT (?product)) as ?total) WHERE \{    ?a v:type "review" . ?a e:edge ?product .  \}} & BGP & 2787 & 2787 \\ \hline
F3 & \small{SELECT DISTINCT ?pid WHERE \{ ?a v:productID ?pid . ?a v:ProductPropertyNumeric\_1 ?property1 . FILTER ( ?property1 = 1 ) \}} & FILTER & ?pid  bsbm:inst/Product1636  bsbm:inst/Product2295 & \{ pid=1636 \}  \{ pid=2295 \} \\ \hline
L2 & \small{SELECT ?rating1 WHERE \{ ?a v:type "review" . ?a v:Rating\_1 ?rating. ?a e:edge ?product. ?product v:productID ?pid . FILTER ( ?pid = 343 ) .\} LIMIT 2} & LIMIT & ?rating1 9 7 & \{ rating1=9 \} \{ rating1=7 \} \\ \hline
G2 & \small{SELECT ?product WHERE \{ ?a v:type "reviewer" . ?a v:reviewerID ?rid. ?a e:edge ?review . ?review v:Rating\_1 ?rating1. ?review e:edge ?product. ?product v:productID ?pid. FILTER ( ?rid = 86). \} GROUP BY (?rating1)} & GROUP BY & ?product bsbm:inst/Product1107 bsbm:inst/Product1301 bsbm:inst//Product1852 bsbm:inst/Product2291 bsbm:inst/Product1098 bsbm:inst/Product1954 bsbm:inst/Product1994 bsbm:inst/Product1355 bsbm:inst/Product734 bsbm:inst/Product1448 bsbm:inst/Product1426 bsbm:inst/Product1817 bsbm:inst/Product1141 bsbm:inst/Product1194 bsbm:inst/Product451 bsbm:inst/Product1294 bsbm:inst/Product1532 & \{ product=1107 \}  \{ product=1301 \}  \{ product=1852 \}  \{ product=2291 \}  \{ product=1098 \}  \{ product=1954 \}  \{ product=1994 \}  \{ product=1355 \}  \{ product=734 \}  \{ product=1448 \}  \{ product=1426 \}  \{ product=1817 \}  \{ product=1141 \}  \{ product=1194 \}  \{ product=451 \}  \{ product=1294 \}  \{ product=1532 \} \\ \hline
Gc2 & \small{SELECT ?product (COUNT (?review) as ?total) WHERE \{ ?review v:type "review" . ?review e:edge ?product . ?product v:productID ?pid. \} GROUP BY (?product) LIMIT 10} & GROUP COUNT & ?product  ?total bsbm:inst/Product2588  1 bsbm:inst/Product3  1 bsbm:inst/Product2331  2 bsbm:inst/Product2553  3 bsbm:inst/Product1803 5 bsbm:inst/Product2440  7 bsbm:inst/Product2201 5  bsbm:inst/Product316 3 bsbm:inst/Product2210  7  & \{Product=2588, Total=1\} \{Product=3, Total=1\} \{Product=2331, Total=2\} \{Product=2553, Total=3\} \{ Product=1803,Total=5 \}
\{ Product=2440, Total=7 \} \{ Product=2201, Total=5 \} \{ Product=316, Total=3 \} \{ Product=2210, Total=7 \} \\ \hline
O2 & \small{SELECT DISTINCT ?product ?label WHERE \{  ?a v:productTypeID ?tid. FILTER(?tid = 58). ?a e:edge ?product. ?product v:productID ?pid. ?product v:label\_n ?label. \} ORDER BY (?product) LIMIT 5 } & ORDER BY & product        label bsbm:inst/Product11        "pipers pests" bsbm:inst/Product18        "boondogglers"  bsbm:inst/Product489        "airsickness simplices skiing"  bsbm:inst/Product694       "nahuatls terrifiers direr" bsbm:inst/Product709     "jacinth medusoids" & \{pid=11, lab=pipers pests\} \{pid=18, lab=boondogglers\}  \{pid=489, lab=airsickness simplices skiing\} \{pid=694, lab=nahuatls terrifiers direr\} \{pid=709, lab=jacinth  medusoids\} \\ \hline
U1 & \small{SELECT ?label WHERE \{   \{ ?a v:productTypeID ?tid. FILTER(?tid = 58). ?a e:edge ?product. ?product v:productID ?pid. ?product v:label\_n ?label. \}UNION  \{      ?a v:productTypeID ?tid. FILTER(?tid = 102). ?a e:edge ?product. ?product v:productID ?pid. ?product v:label\_n ?label. \}\} LIMIT 10} & UNION & ?label "airsickness simplices skiing" "nahuatls terrifiers direr" "jacinth medusoids" "slowed cloche" "meshwork" "nonradical warehousing" "furnacing" "accommodator" "collectivized mathematics" "brachiate writeoff"  &  \{ label=airsickness simplices skiing \}   \{ label=nahuatls terrifiers direr \}   \{ label=jacinth medusoids \}   \{ label=slowed cloche \}   \{ label=meshwork \}   \{ label=nonradical \} \{ label=warehousing \}   \{ label=furnacing \}   \{ label=accommodator \}   \{ label=collectivized mathematics \}   \{ label=brachiate writeoff \}  \\ \hline
Op1 & \small{SELECT  ?pTex2 ?pText3 ?pNum2  WHERE \{
 ?product v:productID ?pid . FILTER ( ?pid = 343 ) . ?product rdfs:label ?label.
?product v:ProductPropertyTextual2 ?propertyTextual\_2 . ?product v:ProductPropertyTextual3 ?propertyTextual\_3 . OPTIONAL \{ ?product v:productID ?pid . FILTER ( ?pid = 350 ) . ?product rdfs:label ?label. ?product v:ProductPropertyNumeric\_2 ?propertyNumeric2 . ?product v:ProductPropertyTextual3 ?propertyTextual\_3 .\}\}} & OPT. & pText2        pText3        pNum2 ""cyanided uncharged gametes""  ""fluorosis appeasing railheads  criticizers satirizer controllers"" 758\ & \{pText\_2=cyanided uncharged gametes, pText\_3=fluorosis appeasing railheads criticizers satirizer controllers, pNum2\_2=758\} \\ \hline
M1 & \small{SELECT ?reviewer (COUNT (?product) as ?total) WHERE \{ ?reviewer v:type "reviewer". ?reviewer e:edge ?review. ?review e:edge ?product . \} GROUP BY (?reviewer) ORDER BY DESC (?total) LIMIT 10} & MIX & bsbm:inst/Reviewer1294  42 bsbm:inst/Reviewer501       41  bsbm:inst/Reviewer424    39  bsbm:inst/Reviewer281  38  bsbm:inst/Reviewer1263     38  & {[}1294:42, 501:41, 424:39, 281:38, 1263:38{]} \\ \hline
S1 & \small{SELECT ?plabel ?label ?flabel ?proptext1 ?proptext2 ?proptext3 ?propnum1 ?propnum2 ?comment WHERE \{ ?producer v:type "producer". ?producer v:label\_n ?plabel. ?producer e:edge ?product. ?product v:type "product". ?product v:productID ?pid. FILTER(?pid = 343). ?product v:label\_n ?label. ?product v:comment ?comment. ?product v:ProductPropertyTextual\_1 ?proptext1. ?product v:ProductPropertyTextual\_2 ?proptext2. ?product v:ProductPropertyTextual\_3 ?proptext3. ?product v:ProductPropertyNumeric\_1 ?propnum1. ?product v:ProductPropertyNumeric\_2 ?propnum2. ?product e:edge ?pfeature. ?pfeature v:type "product\_feature". ?pfeature v:label\_n ?flabel. \} LIMIT 1} & STAR & ?label        ?comment        ?p        ?f        ?productFeature        ?producer        ?propertyTextual1        ?propertyTextual2        ?propertyTextual3        ?propertyNumeric\_1        ?propertyNumeric1\_2  "ors"  "sobbers kynurenic undergoing remained horsed sidings hutzpa continence flighty japingly semiretired crispest chukkers bamboozler shivah lagged miggs snickering arbitrators propped osmic mismeeting dissimulate fraudulently cabled yeller truncheons sigil expatriating viceless merrymakers fetas recompenses disreputability taperer multiplexed toddler disaffiliating radiating worshipper flamboyance waggly bothering swindlers eucharistical enserfing lightfaced tench tramping margraves bewilderment deuteronomy contravened fourpenny coveralls traitorousness millpond redetermine jeremiad resealable abreaction marblers whisks"  bsbm:inst/Producer8         bsbm:inst/ProductFeature11         "entoiling"  "assignat disrobe"  "housewifeliness neoliths proselytizers infirmable meditations bedchair maschera hagfish saplings prearranges debacles bedews straying grouter stereophonically"  "cyanided uncharged gametes"  "fluorosis appeasing railheads criticizers satirizer controllers"  1165  1526 & {[ProductPropertyNumeric\_1:[1165], productID:[343], ProductPropertyTextual\_1:[cyanided uncharged gametes], ProductPropertyNumeric\_2:[1526], ProductPropertyTextual\_2:[fluorosis appeasing railheads criticizers satirizer controllers], label\_n:[ors], comment:[sobbers kynurenic undergoing remained horsed sidings hutzpa continence flighty japingly semiretired crispest chukkers bamboozler shivah lagged miggs snickering arbitrators propped osmic mismeeting dissimulate fraudulently cabled yeller truncheons sigil expatriating viceless merrymakers fetas recompenses disreputability taperer multiplexed toddler disaffiliating radiating worshipper flamboyance waggly bothering swindlers eucharistical enserfing lightfaced tench tramping margraves bewilderment deuteronomy contravened fourpenny coveralls traitorousness millpond redetermine jeremiad resealable abreaction marblers whisks],type:[product]],label:hedgehogs barstools,label\_prod:assignat disrobe} \\
\bottomrule 
\end{tabular}%
}
\caption{\textbf{Comparison of results of a subset of SPARQL queries and their corresponding Gremlin traversals for BSBM dataset. The complete list of all the queries and their corresponding results can be accessed from the spreadsheet available at~(\url{https://goo.gl/CSSVzZ})}.}
\label{tab:bsbm_1m_results}
\end{table*} 

\section{{\color{cyan} SPARQL - Gremlin Performance Comparison}}\label{sec:app_det_res}
In this section, we present the query-wise detailed results in tabular format of the same  plots reported previously in Figures~\ref{fig:bsbm_all_results} and~\ref{fig:nw_all_results}. 
\begin{landscape}
\begin{table}[htbp]
\centering
\begin{tabular}{|c|p{0.8cm}p{0.8cm}p{0.8cm}p{0.8cm}p{0.8cm}p{0.8cm}|p{0.8cm}p{0.8cm}p{0.8cm}p{0.8cm}p{0.8cm}p{0.8cm}|p{0.8cm}p{0.8cm}p{0.8cm}p{0.8cm}p{0.8cm}p{0.8cm}|}
\rowcolor[HTML]{C0C0C0} 
{\color[HTML]{963400} \textbf{Query}} & \multicolumn{6}{c}{\cellcolor[HTML]{C0C0C0}{\color[HTML]{963400} \textbf{Gremlin Traversal Execution Time (ms, without indexes)}}} & \multicolumn{6}{c}{\cellcolor[HTML]{C0C0C0}{\color[HTML]{963400} \textbf{SPARQL Query Execution Time (ms, with indexes)}}} & \multicolumn{6}{c}{\cellcolor[HTML]{C0C0C0}{\color[HTML]{963400} \textbf{Gremlin Traversal Execution Time (ms, with indexes)}}} \\
\textbf{} & \textbf{Tinker (c)} & \textbf{Tinker (w)} & \textbf{Neo4J (c)} & \textbf{Neo4J (w)} & \textbf{Spaksee (c)} & \textbf{Spaksee (w)} & \textbf{Virtuoso (c)} & \textbf{Virtuoso (w)} & \textbf{Jena (c)} & \textbf{Jena (w)} & \textbf{4store (c)} & \textbf{4store (w)} & \textbf{Tinker (c)} & \textbf{Tinker (w)} & \textbf{Neo4J (c)} & \textbf{Neo4J (w)} & \textbf{Spaksee (c)} & \textbf{Spaksee (w)} \\ \hline
\textbf{C1} & 220.12 & 136.7 & 177.60 & 152.72 & 306.8 & 272.67 & 458.15 & 167.25 & 1652 & 1614 & 260 & 152.5 & 191.6 & 107.69 & 157.3 & 80.05 & 253.82 & 184 \\
\textbf{C2} & 302.6 & 187.3 & 272.67 & 231.36 & 340.84 & 224.48 & 61.25 & 21.5 & 390 & 361.25 & 232.5 & 65 & 238.1 & 100.52 & 218.46 & 97.14 & 279.01 & 203.7 \\
\textbf{C3} & 17.7 & 15.4 & 18.50 & 15.29 & 38.86 & 13.81 & 271 & 70.25 & 404.63 & 384 & 227.5 & 70 & 1.5 & 0.33 & 9.43 & 2.86 & 29.5 & 3.68 \\
\textbf{F1} & 16.32 & 15.4 & 19.72 & 16.40 & 32.66 & 22.08 & 172.25 & 45.5 & 448 & 417.5 & 275.25 & 72.5 & 17.83 & 15.27 & 11.93 & 4.75 & 26.2 & 3.49 \\
\textbf{F2} & 45.6 & 33.5 & 67.06 & 53.17 & 102.61 & 73.70 & 84.15 & 23.75 & 945.5 & 871.5 & 632 & 148.25 & 1.12 & 0.9 & 34.67 & 16.3 & 49.7 & 23.82 \\
\textbf{F3} & 22.8 & 20.1 & 29.37 & 25.81 & 44.93 & 25.34 & 43.3 & 15 & 812.75 & 904.3 & 655 & 132.5 & 1.15 & 0.72 & 2.4 & 1.15 & 29 & 3.77 \\
\textbf{L1} & 17.6 & 16 & 34.73 & 33.09 & 39.20 & 19.97 & 14.45 & 7.25 & 379.15 & 369.25 & 245 & 72.5 & 0.246 & 0.169 & 3.85 & 1.74 & 17.8 & 5.3 \\
\textbf{L2} & 107.17 & 56.57 & 71.37 & 32.05 & 105.26 & 68.42 & 44.5 & 21.25 & 456.75 & 416.5 & 287.5 & 55.3 & 77.9 & 49.41 & 45.26 & 17.98 & 43.8 & 21.52 \\
\textbf{L3} & 6.84 & 6.4 & 18.80 & 14.77 & 9.09 & 8.53 & 8.3 & 4.75 & 433 & 429.75 & 382.25 & 175 & 0.79 & 0.63 & 3.25 & 1.72 & 12.75 & 4.29 \\
\textbf{G1} & 42.57 & 28.87 & 50.14 & 22.49 & 48.11 & 43.54 & 61.45 & 22.5 & 811.34 & 544.21 & 465.19 & 132.75 & 18.52 & 8.01 & 28.3 & 4.18 & 29.3 & 9.29 \\
\textbf{G2} & 19.15 & 16.23 & 35.70 & 16.58 & 20.76 & 18.87 & 272.3 & 69.25 & 434.3 & 383.75 & 212.5 & 60.34 & 18.5 & 16.5 & 18.2 & 3.98 & 11.67 & 3.89 \\
\textbf{G3} & 260.47 & 149.1 & 248.53 & 156.58 & 306.06 & 287.54 & 77.25 & 50.5 & 916.5 & 858 & 682.5 & 130 & 269.5 & 163.3 & 133.4 & 54.52 & 212.82 & 174.25 \\
\textbf{Gc1} & 24.6 & 22.36 & 15.85 & 11.27 & 29.59 & 25.2 & 289 & 171.25 & 1535 & 1450 & 425 & 305 & 0.33 & 0.268 & 4.92 & 2.08 & 7.92 & 3.05 \\
\textbf{Gc2} & 23.47 & 21.36 & 21.49 & 17.09 & 27.04 & 24.85 & 110.3 & 94.25 & 2809.75 & 2717.3 & 2727.5 & 1110.25 & 0.737 & 0.679 & 6.07 & 2.34 & 5.88 & 3 \\
\textbf{Gc3} & 23.98 & 21.6 & 19.27 & 15.62 & 28.78 & 25.01 & 102 & 30 & 1063 & 1045.75 & 287.5 & 120.75 & 0.762 & 0.667 & 5.19 & 2.25 & 6.31 & 2.79 \\
\textbf{O1} & 328.57 & 192.77 & 238.3 & 173.95 & 461.59 & 402.64 & 183.5 & 164.7 & 1234.25 & 1200 & 207.45 & 130.5 & 294.68 & 193.44 & 210.2 & 77.75 & 331.9 & 227.24 \\
\textbf{O2} & 20 & 17.3 & 32.93 & 25.32 & 56.64 & 35.68 & 44.5 & 15.3 & 530 & 403.25 & 278 & 87.5 & 4.56 & 2.63 & 8.12 & 3.96 & 26.7 & 3.39 \\
\textbf{O3} & 525.81 & 357.98 & 489.63 & 369.2 & 612.04 & 483.51 & 604 & 519 & 2645.5 & 2109 & 650.9 & 321.65 & 205.1 & 158.18 & 196.87 & 153.76 & 338.74 & 179.82 \\
\textbf{U1} & 45.72 & 23.17 & 53.15 & 48.6 & 111.62 & 58.78 & 44.89 & 13.25 & 498.5 & 504 & 207.5 & 90 & 18.98 & 9.42 & 28.51 & 13.49 & 47.5 & 15.59 \\
\textbf{U2} & 373.95 & 203.61 & 402.40 & 378.1 & 503 & 286.24 & 34 & 10.75 & 444.5 & 403 & 190 & 55 & 257.5 & 173.2 & 159.01 & 68.63 & 307.53 & 152.7 \\
\textbf{U3} & 278.89 & 173.01 & 328.60 & 287.29 & 565.19 & 329.1 & 151.4 & 40.75 & 1111 & 1169 & 507.5 & 190 & 218.73 & 153.9 & 141.52 & 47.14 & 224.1 & 72.67 \\
\textbf{M1} & 1221.7 & 765.7 & 246.25 & 116.57 & 453.82 & 367.93 & 74.15 & 58.5 & 3445.15 & 3417.5 & 1535 & 498 & 899 & 618.25 & 126.33 & 75.79 & 226.23 & 130.7 \\
\textbf{M2} & 806.1 & 537.4 & 283.76 & 161.43 & 327.01 & 276.04 & 120.6 & 98.25 & 3513.75 & 3418.25 & 2857.5 & 1412.5 & 657.6 & 451 & 143.61 & 87.81 & 139.6 & 55.84 \\
\textbf{M3} & 551.6 & 309.32 & 161.72 & 81.63 & 261.92 & 168.91 & 84 & 75.5 & 3384.25 & 3367 & 1080 & 343 & 402.7 & 235.9 & 66.48 & 34.36 & 102.35 & 39.37 \\
\textbf{S1} & 20.75 & 15.5 & 25.38 & 16.72 & 38.83 & 32.79 & 388.15 & 108.5 & 593 & 630 & 253 & 70 & 1.55 & 0.79 & 2.54 & 1.97 & 12.91 & 3.94 \\
\textbf{S2} & 17.5 & 15.4 & 21.16 & 14.8 & 26.45 & 21.89 & 332 & 89.5 & 442 & 413 & 202.5 & 80 & 1.45 & 0.82 & 2.39 & 1.71 & 10.18 & 3.42 \\
\textbf{S3} & 34.02 & 26.65 & 40.75 & 31.54 & 52.57 & 40.2 & 40641 & 40329 & 22663 & 21180 & 2175 & 1730 & 7.02 & 4.46 & 15.92 & 8.02 & 31.2 & 6.84 \\ \bottomrule
\end{tabular}
\caption{\textbf{Performance comparison of SPARQL queries vs translated Gremlin traversals over BSBM dataset.}}
\label{tab:bsbm_full_stats}
\end{table}
\end{landscape}


\begin{landscape}
\begin{table}[htbp]
\centering
\begin{tabular}{|c|p{0.8cm}p{0.8cm}p{0.8cm}p{0.8cm}p{0.8cm}p{0.8cm}|p{0.8cm}p{0.8cm}p{0.8cm}p{0.8cm}p{0.8cm}p{0.8cm}|p{0.8cm}p{0.8cm}p{0.8cm}p{0.8cm}p{0.8cm}p{0.8cm}|}
\rowcolor[HTML]{C0C0C0} 
{\color[HTML]{963400} \textbf{Query}} & \multicolumn{6}{c}{\cellcolor[HTML]{C0C0C0}{\color[HTML]{963400} \textbf{Gremlin Traversal Execution Time (ms, without indexes)}}} & \multicolumn{6}{c}{\cellcolor[HTML]{C0C0C0}{\color[HTML]{963400} \textbf{SPARQL Query Execution Time (ms, with indexes)}}} & \multicolumn{6}{c}{\cellcolor[HTML]{C0C0C0}{\color[HTML]{963400} \textbf{Gremlin Traversal Execution Time (ms, with indexes)}}} \\
\textbf{} & \textbf{Tinker (c)} & \textbf{Tinker (w)} & \textbf{Neo4J (c)} & \textbf{Neo4J (w)} & \textbf{Spaksee (c)} & \textbf{Spaksee (w)} & \textbf{Virtuoso (c)} & \textbf{Virtuoso (w)} & \textbf{Jena (c)} & \textbf{Jena (w)} & \textbf{4store (c)} & \textbf{4store (w)} & \textbf{Tinker (c)} & \textbf{Tinker (w)} & \textbf{Neo4J (c)} & \textbf{Neo4J (w)} & \textbf{Spaksee (c)} & \textbf{Spaksee (w)} \\ \hline
\textbf{C1} & 11.49 & 5.45 & 17.7 & 13.3 & 23.89 & 13.8 & 36.25 & 3.75 & 364 & 374 & 298 & 93 & 7.75 & 4.25 & 12.87 & 6.17 & 14.4 & 6.64 \\
\textbf{C2} & 20.757 & 11.8 & 28.3 & 21.2 & 31.13 & 15.7 & 25 & 2.75 & 395 & 389 & 290 & 80 & 11.52 & 6.14 & 18.4 & 5.26 & 23.9 & 10.52 \\
\textbf{C3} & 10.76 & 5.84 & 18.9 & 12.6 & 24 & 11.4 & 52 & 4.25 & 399 & 392 & 318 & 90 & 7.5 & 4.25 & 12.38 & 4.58 & 15.17 & 7.1 \\
\textbf{F1} & 0.77 & 0.8 & 3.49 & 2.7 & 4.45 & 2.4 & 12 & 4.33 & 363 & 369 & 273 & 115 & 0.53 & 0.29 & 2.16 & 1.03 & 2.45 & 1.32 \\
\textbf{F2} & 15.45 & 8.47 & 24.5 & 12.09 & 34.3 & 17.62 & 6 & 2 & 370 & 427 & 270 & 124 & 9.6 & 6.25 & 19.52 & 9.26 & 24.4 & 11.08 \\
\textbf{F3} & 2.4 & 1.24 & 6.68 & 5.41 & 11.28 & 9.7 & 5 & 2.65 & 470 & 373 & 252 & 115 & 1 & 0.72 & 8.15 & 3.38 & 7.6 & 3.46 \\
\textbf{L1} & 1.5 & 0.73 & 4.49 & 3.6 & 5.3 & 4.09 & 7.75 & 5 & 439 & 349 & 280 & 112 & 0.55 & 0.4 & 3.01 & 1.39 & 4.5 & 2.16 \\
\textbf{L2} & 2.96 & 0.47 & 7.43 & 4.6 & 4.75 & 2.8 & 12 & 2.5 & 364 & 364 & 295 & 113 & 0.39 & 0.38 & 4.12 & 1.83 & 4.57 & 2.25 \\
\textbf{L3} & 17.01 & 8.98 & 29.1 & 18.42 & 37.5 & 24.2 & 4 & 2.5 & 410 & 378 & 279 & 108 & 11.43 & 6.42 & 21.68 & 8.09 & 27.1 & 15.32 \\
\textbf{G1} & 6 & 2.15 & 12.16 & 8.81 & 4.39 & 3.2 & 64 & 18.25 & 433 & 461 & 245 & 103 & 1.46 & 1.11 & 3.22 & 1.26 & 3.04 & 2.27 \\
\textbf{G2} & 5.14 & 1.8 & 10.6 & 5.5 & 3.62 & 2.7 & 27 & 9.75 & 348 & 369 & 285 & 92 & 2.29 & 0.9 & 3.72 & 2.35 & 2.8 & 2.19 \\
\textbf{G3} & 6.32 & 1.5 & 13.27 & 9.76 & 4.86 & 3.8 & 2.75 & 2 & 471 & 476 & 328 & 157 & 1.34 & 0.85 & 5.25 & 3.99 & 3.95 & 3.04 \\
\textbf{Gc1} & 10.2 & 5.45 & 17.3 & 13.6 & 15.49 & 9.52 & 9 & 6.33 & 440 & 467 & 247 & 88 & 7.25 & 3.9 & 10.83 & 7.05 & 9.66 & 4.6 \\
\textbf{Gc2} & 25 & 14.09 & 38.1 & 23.9 & 36.6 & 15.67 & 6 & 2.75 & 385 & 354 & 250 & 105 & 15.92 & 9.25 & 23.62 & 9.84 & 18.9 & 7.01 \\
\textbf{Gc3} & 0.783 & 0.68 & 6.89 & 3.7 & 4.61 & 3.8 & 47.75 & 4.25 & 753 & 775 & 248 & 118 & 0.52 & 0.47 & 3.42 & 3.33 & 3.52 & 3.43 \\
\textbf{O1} & 18.25 & 12.23 & 28 & 17.4 & 40.6 & 25.3 & 4.5 & 2.25 & 446 & 483 & 267 & 125 & 12.32 & 7.5 & 19.1 & 8.44 & 27.46 & 12.2 \\
\textbf{O2} & 28 & 17.4 & 39.2 & 23 & 54.3 & 27.81 & 20 & 8.25 & 321 & 354 & 270 & 95 & 19.2 & 11.93 & 22.3 & 6.85 & 35.49 & 13.8 \\
\textbf{O3} & 18.45 & 10.44 & 28.3 & 18.02 & 42.78 & 24.64 & 3.75 & 2.25 & 545 & 573 & 300 & 120 & 10.65 & 6.6 & 18.4 & 5.41 & 28.27 & 11.3 \\
\textbf{U1} & 67.25 & 36.75 & 70.6 & 46.25 & 81.72 & 39.78 & 27.75 & 3.75 & 377 & 368 & 240 & 115 & 43.75 & 24.11 & 35.64 & 15.46 & 45.34 & 28.76 \\
\textbf{U2} & 68.5 & 35.45 & 89.8 & 42.61 & 107.76 & 48.28 & 16.5 & 2.75 & 529 & 536 & 252.5 & 130 & 40 & 23.25 & 53.21 & 29.71 & 69.2 & 33.71 \\
\textbf{U3} & 75.5 & 42.6 & 98.5 & 54.4 & 157.6 & 62.91 & 31.5 & 4.5 & 346 & 434 & 260 & 112.5 & 44.5 & 23.58 & 65.31 & 28.33 & 120.49 & 73.79 \\
\textbf{M1} & 22.45 & 12.4 & 37.8 & 23.4 & 49.14 & 24.5 & 12 & 6.25 & 457 & 473 & 325 & 140 & 12.87 & 6.9 & 18.09 & 7.97 & 24.1 & 13.8 \\
\textbf{M2} & 2.56 & 1.65 & 6.94 & 3.67 & 51.2 & 26.3 & 8 & 3 & 373 & 423 & 301 & 118 & 1.6 & 1.13 & 6.02 & 3.26 & 28.6 & 15.36 \\
\textbf{M3} & 2.5 & 1.92 & 7 & 4.16 & 50.8 & 25.4 & 35 & 14.5 & 563 & 547 & 298 & 110 & 1.8 & 1.25 & 4.07 & 3.38 & 27.8 & 15.59 \\
\textbf{S1} & 35.8 & 28.4 & 42.6 & 20.9 & 35.6 & 15.9 & 176 & 67.5 & 451 & 372 & 278 & 135 & 32.5 & 17.3 & 26.47 & 12.15 & 21.4 & 10.63 \\
\textbf{S2} & 7.85 & 5.03 & 11.8 & 5.3 & 21.4 & 10.8 & 50 & 24.75 & 328 & 369 & 253 & 98 & 4.09 & 3.25 & 2.25 & 2.02 & 18.9 & 8.51 \\
\textbf{S3} & 4.23 & 1.42 & 6.9 & 3.8 & 17.6 & 8.3 & 902 & 824 & 362 & 406 & 250 & 80 & 1.25 & 0.82 & 9.15 & 5.01 & 11.9 & 6.02 \\ \hline
\end{tabular}
\caption{\textbf{Performance comparison of SPARQL queries vs translated Gremlin traversals over Northwind dataset.}}
\label{tab:nw_full_stats}
\end{table}
\end{landscape}

\end{appendix}



\nocite{*} 
\bibliographystyle{ios1}           
\bibliography{ref}        

%

\end{document}